\documentclass{aa}  
\usepackage{graphicx}
\usepackage{subfigure}
\usepackage{txfonts}
\usepackage{natbib}
\usepackage{rotating}
\usepackage{longtable,lscape}
\bibpunct{(}{)}{;}{a}{}{,}
\begin{document}
 \title{Detection of vibrationally excited methyl formate in W51 e2}

   \author{K. Demyk \inst{1}\thanks{\emph{Present address: Centre d'Etude Spatiale des Rayonnements, Universit\'e Paul Sabatier, 
                    F-31028 Toulouse, France }} 
          \and
          G. Wlodarczak \inst{1}
          \and
          M. Carvajal \inst{2}
          }

  \offprints{K. Demyk, \email{karine.demyk@cesr.fr}}

   \institute{Laboratoire de Physique des Lasers, Atomes et Mol\'ecules, UMR CNRS 8523
              Universit\'e Lille 1, F-59655 Villeneuve d'Ascq Cedex\\
              \email{Karine.demyk@cesr.fr}
   \and       
              Departamento de Fisica Aplicada, 
              Facultad de Ciencias Experimentales, Universidad de Huelva, 
21071 Huelva, Spain
           }

   \date{Received 7 January 2008; accepted 30 June  2008}

  \abstract
   {Hot cores in molecular clouds, such as Orion KL, Sgr B2, W51 e1/e2, are characterized by the presence of molecules at temperature high enough to significantly populate their low-frequency vibrationally excited states. For complex organic molecules, characterized by a dense spectrum both in the ground state and in the excited states, such as methyl formate, ethyl cyanide or dimethyl ether, lines from vibrationally excited states certainly participate to the spectral confusion.}
   {Thanks to the recent laboratory study of the first torsional excited mode of methyl formate, we have searched for methyl formate, HCOOCH$_3$, in its first torsionally excited state ($\upsilon_t$=1) in the molecular cloud W51 e2.}
   {We have performed observations of the molecular cloud W51 e2 in different spectral regions at 1.3, 2 and 3 mm with the IRAM 30 m single dish antenna.  }
   { Methyl formate in its first torsionally excited state ($\upsilon_t$=1 at 131 cm$^{-1}$) is detected for the first time toward W51 e2. 82 transitions have been detected among which 46 are unblended with other species. For a total of 16 A-E pairs in the observed spectrum, 9 are unblended; these 9 pairs are all detected. All  transitions from excited methyl formate within the observed spectral range are actually detected and no strong lines are missing. The column density of the excited state is comparable to that of the ground state. For a source size of 7\arcsec we find that $\mathrm{T_{rot}}$ = 104 $\pm$ 14 K and N =  9.4$^{+4.0}_{-2.8}$ $\times$10$^{16}$ cm$^{-2}$ for the excited state and  $\mathrm{T_{rot}}$ = 176 $\pm$ 24 K and N =  1.7$^{+.2}_{-.2}$ $\times$ 10$^{17}$ cm$^{-2}$ for the ground state. Lines from ethyl cyanide in its two first excited states ($\upsilon_t$=1, torsion mode at 212 cm$^{-1}$) and ($\upsilon_b$=1, CCN in-plane bending mode at 206 cm$^{-1}$) are also present in the observed spectrum. However blending problems prevent a precise estimate of its abundance.  However as for methyl formate it should be comparable with the ground state for which we find $\mathrm{T_{rot}}$ = 103 $\pm$ 9 K and N =  3.7$^{+0.6}_{-0.4}$ $\times$10$^{15}$ cm$^{-2}$ for a 7\arcsec source size.}
   {With regard to the number of lines of excited methyl formate and ethyl cyanide detected in W51 e2, it appears that excited states of large molecules certainly account for a large number of unidentified lines in spectral survey of molecular clouds. }


\maketitle

\section{Introduction} 

W51 e2 is a hot core part of the W51 H$\mathrm{_{II}}$ region located in the Sagittarius spiral arm at a distance of about 7-8 kpc. It is a region of high-mass star formation.
W51 e2 and W51 e1 are thought to be important star-forming cores. They exhibit a rich chemistry, comparable to that observed in Orion or in Sagittarius. Numerous large organic molecules have been observed towards them. 
CH$_3$CN and CS maps were studied by \citet{zhang98}. Formic acid (HCOOH) was mapped by \cite{liu2001}. Methyl formate (HCOOCH$_3$) and ethyl cyanide (CH$_3$CH$_2$CN) were observed in several studies \citep{liu2001,ikeda2001,remijan2002}. \citet{ikeda2001} studied ethylene oxide (c-C$_2$H$_4$O) and its isomer acetaldehyde (CH$_3$CHO).
Acetic acid (CH$_3$COOH) was detected by \citet{remijan2002} with a fractional abundance of (1-6) $\times$ 10$^{-2}$ relative to HCOOCH$_3$. Glycine, whose presence may be suggested from the observations of acetic acid with which it shares common structural elements, was not detected in W51 e2 \citep{snyder2005}. Recently, trans-ethyl methyl ether was detected in W51 e2 \citep{fuchs05}.

The rotational temperature of most of these large molecular species is high. \citet{liu2001} estimated the rotational temperature in W51 e2 to be in the 200-300 K range. 
From high resolution observations of CH$_3$CN analyzed with statistical equilibrium models, \citet{remijan2004} derived the kinetic temperature in W51 e2 to be T$_{\mathrm{kin}}$ =  153(21) K. At such temperature, low-energy vibrational excited states can be significantly populated. Transitions from vibrationally excited states have indeed been observed in other sources such as in Sgr B2(N-LMH) for C$_2$H$_3$CN, CH$_3$CH$_2$OH \citep{numm98} and CH$_3$CH$_2$CN \citep{mehringer2004}. Recently, lines from torsionally excited methyl formate have been identified in Orion KL \citep{koba2007}.

In this study we present the first detection of excited methyl formate and ethyl cyanide in the molecular cloud W51 e2. The initial project was to look for methyl carbamate (NH$_2$COOCH$_3$), an isomer of glycine (NH$_2$CH$_2$COOH) which has a larger dipole moment, making its detection more favorable than glycine. It was not detected. However, few strong unidentified lines in the data attracted our attention and were possibly attributed to methyl formate in its first torsional excited state. Further observations confirmed this detection and also lead to the detection of excited ethyl cyanide in W51 e2. The first vibrationally excited state of methyl formate is the CH$_3$ torsion mode, $\nu_{18}$, hereafter called $\upsilon_t$=1, at 131 cm$^{-1}$ (188 K). The rotational spectrum in this excited state was measured and analyzed by \citet{ogata2004}. Ethyl cyanide has two close vibrationally excited states, the CCN in-plane bending mode, $\nu_{13}$, hereafter called $\upsilon_b$=1, at 206 cm$^{-1}$  (296 K) and the CH$_3$ torsion mode, $\nu_{21}$, hereafter called $\upsilon_t$=1, at 212 cm$^{-1}$ (305 K). A preliminary analysis of the rotational spectrum in these two excited states is presented in the paper from \citet{mehringer2004}.

The observations and the methods used for the data analysis are described in Sect.~\ref{obs} and Sect.~\ref{analysis}, respectively. The study of methyl formate and ethyl cyanide in the ground and excited state is presented in Sect.~\ref{mtf} and Sect.~\ref{propio}, respectively. The search for methyl carbamate and glycine is presented in Sect. ~\ref{mtc}. The discussion is carried out in Sect.~\ref{discussion}.

\section{Observations}
\label{obs}
The observations were performed with the IRAM 30m antenna at Pico Veleta (Spain) in June 2003 and June 2006. W51 e2 was observed at the position $\alpha$(2000) = 19\degr23\arcmin43.9\arcsec and $\delta$(2000) = 14\degr30\arcmin34.8\arcsec in position switching mode with the OFF position located at $\alpha$ = 300~\arcsec and $\delta$ = 0~\arcsec.

Several spectral windows in the 80-250 GHz range were observed in order to include as many transitions as possible for the searched molecules (excited methyl formate and ethyl cyanide for the 2006 data and methyl carbamate and glycine for the 2003 data) and as few transitions as possible of other molecules having numerous strong lines (such as methyl formate and ethyl cyanide in the ground state, dimethyl ether, etc.).

All lines were observed with an array of 4 receivers (in single side band mode) set at the appropriate frequencies. The spectrometers used are a low resolution 1 MHz filter bank and an autocorrelator with a spectral resolution in the 40-320 kHz range, split between different receivers. Focus and pointing were regularly checked on the nearby ultra compact HII region K 3-50A. The rejection of the image band (USB) was of about 26 dB at 3 mm, 12 dB at 2 mm, 15 dB at 1.3 mm and 10 dB at 1.1 mm. The system temperature was typically 100-200 K, 200-700 K, 200-700 K and 400-1500 K at 3, 2, 1.3 and 1.1 mm, respectively. The total usable ON+OFF integration time varies from 30 to 50 minutes depending on the frequency range. The beam size is 22~\arcsec, 17~\arcsec and 10.5~\arcsec at 3, 2 and 1.3 mm, respectively. The spectra are presented in main beam temperature unit which is calculated from the antenna temperature: T$_\mathrm{mb}$ = F$_\mathrm{eff}$/B$_\mathrm{eff}$ $\times$ $\mathrm{T_A}^*$. The data were reduced using the GILDAS package.

\section{Analysis}
\label{analysis}

For the data analysis we assume that local thermodynamic equilibrium (LTE) is reached, i.e. we assume that the excitation, rotational and vibrational temperatures are equal to the kinetic temperature in the emitting region and that the lines are thermalized, i.e. their level population is described by a Boltzmann distribution at that temperature. The validity of this assumption will be discussed later in Sect.~\ref{discussion}.\\

The data were analyzed using the classical rotational diagram method to estimate the rotational temperature and the column density together with their uncertainties for the different identified species. We adopt the formulation from \citet{turner1991}, corrected for beam dilution effects: 

\begin{equation}\label{DR}
\mathrm{ln(\frac{3kW}{8{\pi}^3{\nu}S{\mu}^2g_ig_k}) = ln(\frac{N}{Q}) - \frac{E_u}{kT} - ln(b)}
\end{equation}

\noindent where W is the integrated line intensity in K.km.s$^{-1}$, $\nu$ the line frequency, S${\mu}^2$ the line strength in Debye$^2$, $g_i$ the reduced nuclear spin statistical weight, $g_k$ the K-level degeneracy, Q is the partition function, $\mathrm{E_u}$ the upper state energy, N is the total column density and T the excitation temperature. Assuming a gaussian beam, the beam dilution factor b is given by: 

\begin{equation}\label{bf}
\mathrm{b = \frac{\theta_s^2} {\theta_s^2 + \theta_{tel}^2}}
\end{equation}

\noindent where $\mathrm{\theta_s}$ and $\mathrm{\theta_{tel}}$ are, respectively, the source and telescope beam size in arcsecond.\\
Beam dilution effects were taken into account both in the rotational diagram analysis and in the emission modeling (see below). The emission region in W51 e2 is observed to be smaller than 10\arcsec for most organic molecules \citep{remijan2002,liu2001}. Consequently beam dilution effects are important at low frequency at which the IRAM 30m antenna beam size is significantly larger (29\arcsec at 86 GHz). \\

We have compared the observed spectrum with simulated spectra calculated using a simple emission model at local thermodynamic equilibrium (LTE). The expression for the simulated main beam temperature for one molecule is thus: 

\begin{equation}\label{Tmb}
\mathrm{T_{mb} = b \times (J - J_{bg}) \times (1 - e^{-\tau})} 
\end{equation}

\noindent where J is the source function: 

\begin{equation}\label{J}
\mathrm{J = \frac{h\nu}{k} \times (e^{h\nu/kT} -1)^{-1}}
\end{equation}

\noindent and 

\begin{equation}\label{Jbg}
\mathrm{J_{bg} = \frac{h\nu}{k} \times (e^{h\nu/k\times 2.7} -1)^{-1}}
\end{equation}

\noindent $\tau$ is the optical depth, summed over all the transitions of the molecules:

\begin{equation}\label{tau}
\mathrm{\tau = \sum_i \frac{c^2}{8\pi\nu^2}N_{tot}\frac{g_u}{Q}A_{ul}\Phi(\nu)e^{-E_l/kT}(1-e^{-E_{ul}/kT}})
\end{equation}

\noindent and  $\Phi(\nu)$ is the line profile: 

\begin{equation}\label{phi}
\mathrm{\Phi(\nu) = \frac{1}{\sqrt{\pi\Delta\nu_D}} \times e^{-(\nu-\nu_i)^2/\Delta\nu_D^2}}
\end{equation}

\noindent $\mathrm{A_{ul}}$ is the Einstein coefficient, $\mathrm{E_{ul}}$ the energy of the transition, $\mathrm{E_l}$ the energy of the lower state, $\mathrm{g_u}$ is the upper state degeneracy, Q the partition function, $\mathrm{\Delta\nu_D}$ is the Doppler width of the line, b is the beam dilution correction factor and $\mathrm{N_{tot}}$ is the total column density. \\

In hot cores, the temperature is such that the low-energy vibrational and/or torsional excited modes are significantly populated. Consequently we have used the vibrational-rotational partition function, Q$\mathrm{_{rv}}$, instead of the pure rotational partition function. Assuming non-interacting harmonic vibrational levels and rigid rotor levels, the ro-vibrational partition function is approximated by \citep[see][]{gordy}:

\begin{equation}\label{qrv}
\mathrm{Q_{rv} = \prod_i(1-e^{-h{\nu}_i/kT})^{-d_i} \times Q_{rot}}
\end{equation}

\noindent where ${\nu}_i$ is the frequency of the vibrational mode i, $\mathrm{d_i}$ its degeneracy, and Q$\mathrm{_{rot}}$ is the rotational partition function.  Q$\mathrm{_{rot}}$ is approximated by:

\begin{equation}\label{qr}
\mathrm{Q_{rot}} = \sigma  \times \sqrt{\mathrm{\frac{\pi(kT)^3}{h^3ABC}}}
\end{equation}

\noindent where $\mathrm{\sigma}$ is the symmetry number \citep[see][]{gordy}. At LTE, the temperature used to calculate the rotational and vibrational partition function is the same, we thus implicitly assume that T$\mathrm{_{vib}}$ = T$\mathrm{_{rot}}$, an hypothesis which may not be valid (see Sect.\ref{discussion}).  For methyl formate which has one internal rotor, rotational transitions are split into A and E components and $\mathrm{\sigma}$ is equal to  2. To calculate the partition function, we have considered the first excited state of methyl formate at 131 cm$^{-1}$. For ethyl cyanide, we have used the partition function given by \citet{mehringer2004} which is calculated by summing the rotational states of the ground state and of the first two excited states of ethyl cyanide at 206 and 212 cm$^{-1}$. For methyl carbamate we have used the partition function in the ground vibrational state given by \cite{groner07}.

\section{Torsionally excited methyl formate}
\label{mtf}

The prediction of the methyl formate lines is based on the most recent work on this species from \citet{carvajal07}. In this work, all experimental data available on the ground and excited states (3496 and 774 microwave lines, respectively) in the  7-200 GHz frequency range, covering the J values up to 43 in the  ground state and up to 18 in the first excited state $\upsilon_t$=1, were collected from previously published studies \citep{ogata2004,oesterling1999,plummer1986, plummer1984,demaison1984}.  \citet{carvajal07} also added 434 new lines of methyl formate in the ground state, measured in Lille in the 567-669 GHz spectral range and corresponding to transitions with J and K values up to 62 and 22, respectively. This dataset was fitted within almost experimental accuracy (root-mean-square deviations of 94 kHz and 84 kHz for the 3496 (774) lines of the ground torsional state and of the excited state $\upsilon_t$=1, respectively) using the so-called ÔÔrho axis methodÕÕ (RAM) described in the literature \citep{hougen94} and a model extended to include perturbation terms through eighth order. The spectroscopic parameters and the details on the fitting procedure are given in the paper from \citet{carvajal07} in which a table presenting all the fitted experimental frequencies, measurement uncertainties, calculated frequencies, observed-calculated values , line strengths, energy levels as well as identification of the transitions, is available as Supplementary data. 

For the detection in the W51 e2 spectrum, we have provided a line-list of predicted line-center frequencies and line intensities based on an internal rotation model (RAM or Rho Axis Method). This method and the code\footnote{A version of the program is available at the web site (http://www.ifpan.edu.pl/$\mathrm{\sim}$kisiel/introt/introt.htm$\mathrm{\sharp}$belgi) and other versions can be available by I. Kleiner (see the web site for more information).} developed was used for several molecules detected in the interstellar medium  (acetaldehyde, CH$_3$COH, \citep{kleiner96}, acetamide CH$_3$CONH$_2$, \citep{hollis06} and acetic acid CH$_3$COOH, \citep{ilyushin07}. The laboratory measurements and predicted line frequencies of transitions in the first excited torsional state $\upsilon_t$=1 of methyl formate in the spectral range used in the present detection are presented in Table~\ref{mtfvt1lab} together with the line assignment, the observed-calculated value, the experimental accuracy, the calculated uncertainty, the line strength and the energy of the lower level. \footnote{A prediction of the overall spectrum of methyl formate in the ground and first excited states, on a large frequency range will be published soon (Kleiner et al. in preparation).} \\

\addtocounter{table}{1}
\addtocounter{table}{2}

\begin{figure*}
    \centering
    \includegraphics[width=\textwidth]{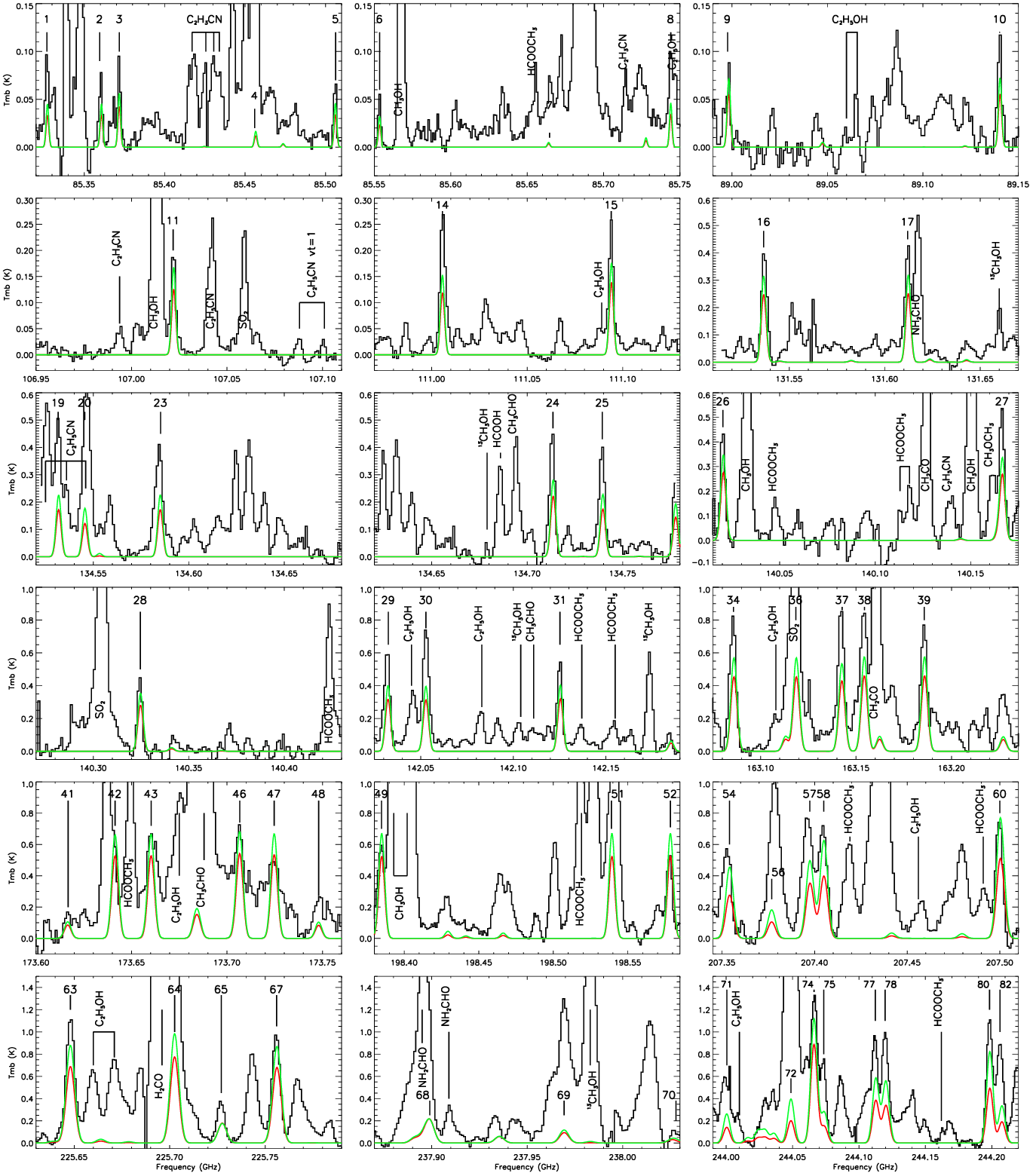}%
    \caption{Detected lines from the first torsionally excited state of methyl formate HCOOCH$_3$. The observations (histogram like curve) are compared with LTE emission models of HCOOCH$_3$ $\upsilon_t$ = 1with ${\theta}_s$ = 7\arcsec, $\mathrm{T_{rot}}$ = 104 K  and N =  9.4 $\times$10$^{16}$ cm$^{-2}$ (red curve) and with ${\theta}_s$ = 12", $\mathrm{T_{rot}}$ = 154 K and N =  5.6 $\times$ 10$^{16}$ cm$^{-2}$ (green curve). The numbered lines are transitions from torsionally excited HCOOCH$_3$, see Table~\ref{mtfvt1} for the attribution of each line.}%
    \label{mtfvt1lte}
\end{figure*}

More than eighty transitions from torsionally excited methyl formate, 82 precisely, are detected in the source among which 46 are not blended.
Taking into account possible blending, we find that all lines from torsionally excited methyl formate predicted to be intense enough to be detectable are present in the observed spectra. Furthermore, no lines, such as unobserved  strong lines, contradicts the identification in the observed spectral range. The internal rotation of the methyl group of methyl formate  splits the transitions into A and E components having the same intensity. Among the 16 observed A-E pairs, 9 are not blended (see  Fig.~\ref{mtfvt1lte}). The intensity of the lines in a A-E pair, when none of the lines in the pair is blended, is consistent with the expected ones, strengthening the identification of excited methyl formate in this source. The detected lines (observed frequency, integrated intensity and line intensity) are listed in Table~\ref{mtfvt1} together with the laboratory or calculated frequency, the quantum numbers of the transition, its line strength and lower state energy. The first column of Table~\ref{mtfvt1} indicates the line number which is used to label the lines in Fig.~\ref{mtfvt1lte}.  The number in parenthesis points out the A or E line associated to the transition when it is observed. However about half of the lines are blended. Comments have been added in Table~\ref{mtfvt1} to indicate line blenders when they are identified. \\

\begin{figure}
    \includegraphics[width=.5\textwidth]{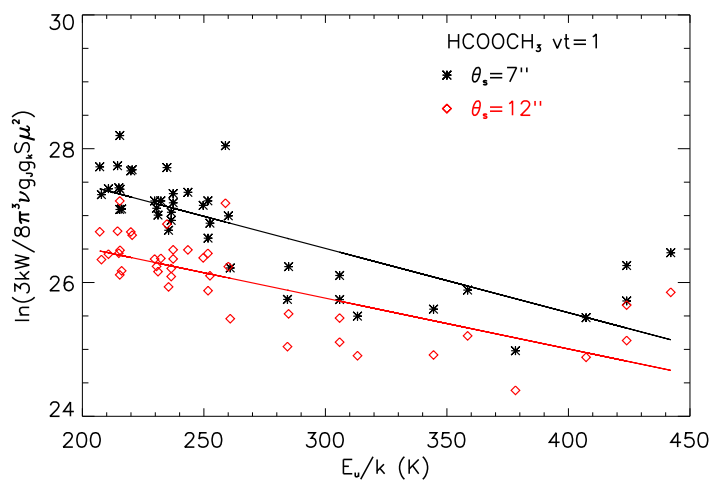}%
    \caption{Rotational diagram of the torsionally excited state of methyl formate for different source sizes. For a source size of 7\arcsec (black stars) we find a rotational temperature of $\mathrm{T_{rot}}$ = 104 $\pm$ 14 K and a column density of N =  9.4$^{+4.0}_{-2.8}$ $\times$10$^{16}$ cm$^{-2}$; for a source size of 12\arcsec (red diamonds) we find $\mathrm{T_{rot}}$ = 131 $\pm$ 20 K and N =  3.4$^{+1.5}_{-1.1}$ $\times$10$^{16}$ cm$^{-2}$.}%
    \label{dr_mtfvt}
\end{figure}

\begin{figure}
    \includegraphics[width=.5\textwidth]{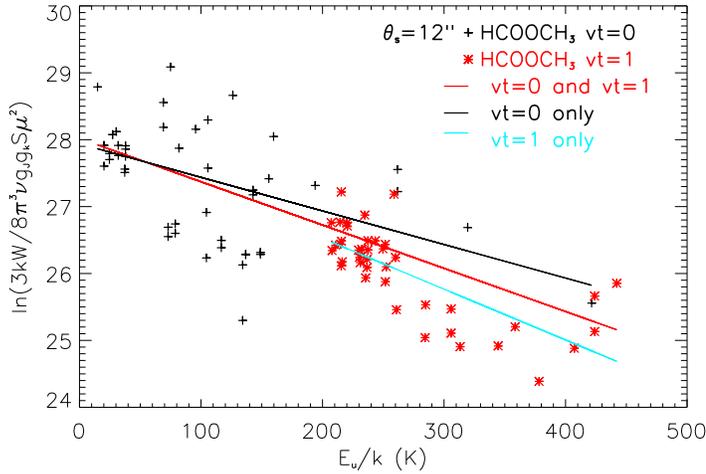}%
    \caption{Rotational diagram of methyl formate in the ground state (black crosses) and of methyl formate both in the ground and excited states (red stars). The source size is 12\arcsec. The rotational temperature and column density for methyl formate in the ground state are $\mathrm{T_{rot}}$ = 199 $\pm$ 28 K and N =  9.6$^{+0.9}_{-0.9}$ $\times$ 10$^{16}$ cm$^{-2}$ and $\mathrm{T_{rot}}$ = 154 $\pm$ 8 K and N =  5.6$^{+4.0}_{-3.7}$ $\times$ 10$^{16}$ cm$^{-2}$ when the ground and the first torsionally excited state are combined. For comparison the rotational diagram for the excited state only is plotted (blue curve).}%
    \label{dr_mtffondvt}
\end{figure}

The non-blended lines have been used to estimate the rotational temperature and column density of excited HCOOCH$_3$ using the rotational diagram method (Fig.\ref{dr_mtfvt},  Table \ref{tcol}). \citet{remijan2002} have mapped W51 e2 with BIMA in two transitions of HCOOCH$_3$ at 228.629 GHz and 90.146 GHz. The size of the emission region in these lines is of about 7\arcsec and 12\arcsec, respectively.
Adopting these values for torsionally excited HCOOCH$_3$, we find a rotational temperature and a column density of $\mathrm{T_{rot}}$ = 131 $\pm$ 20 K and N =  3.4$^{+1.5}_{-1.1}$ $\times$10$^{16}$ cm$^{-2}$ for a source size of 12\arcsec and  $\mathrm{T_{rot}}$ = 104 $\pm$ 14 K and N =  9.4$^{+4.0}_{-2.8}$ $\times$10$^{16}$ cm$^{-2}$ for a source size of 7\arcsec (Table \ref{tcol}). A separate analysis of the A and E lines of excited methyl formate gives compatible values, within the uncertainty, for the rotational temperature and column density. 
Transitions from the ground state of HCOOCH$_3$ have been detected too. Using the rotational diagram method we find for HCOOCH$_3$ $\upsilon_t$=0: $\mathrm{T_{rot}}$ = 199 $\pm$ 28 K and N =  9.6$^{+0.9}_{-0.9}$ $\times$ 10$^{16}$ cm$^{-2}$ for a 12~\arcsec source (Fig. \ref{dr_mtffondvt}).
If we consider both lines from the ground state and from $\upsilon_t$=1 we find $\mathrm{T_{rot}}$ = 154 $\pm$ 8 K and N =  5.6$^{+4.0}_{-3.7}$ $\times$ 10$^{16}$ cm$^{-2}$ for a 12~\arcsec source (Fig. \ref{dr_mtffondvt}, Table \ref{tcol}). These temperatures and column densities have been used to model the emission of methyl formate in the source. The comparison of the modelled spectra with the observations is shown in Fig.~\ref{mtfvt1lte} for a number of lines from HCOOCH$_3$, $\upsilon_t$=1.

\begin{table}
\caption{Temperature and column density of the detected molecules}
\label{tcol}
\centering
\begin{tabular}{ c r  c c}
\hline\hline
Molecules &  & \multicolumn{2}{c}{Source size} \\
	         & & 7\arcsec & 12\arcsec \\
\hline
HCOOCH$_3$ &     T (K)              &176  $\pm$ 24 & 199 $\pm$ 28 \\
$\upsilon_t$=0                &    N (cm$^{-2}$)  & 17.0$^{+1.8}_{-1.6}$ $\times$ 10$^{16}$ & 9.6$^{+0.9}_{-0.9}$ $\times$ 10$^{16}$  \\
\hline
HCOOCH$_3$	&   T (K)              & 104 $\pm$ 14 & 131 $\pm$ 20\\
$\upsilon_t$=1                &   N (cm$^{-2}$)  & 9.4$^{+4.0}_{-2.8}$ $\times$10$^{16}$ &  3.4$^{+1.5}_{-1.1}$ $\times$10$^{16}$\\
\hline
HCOOCH$_3$  & T (K)  &    144 $\pm$ 7   & 154 $\pm$ 8  \\
$\upsilon_t$=0 + $\upsilon_t$=1      &  N (cm$^{-2}$) & 11.4$^{+0.9}_{-0.8}$ $\times$ 10$^{16}$ & 5.6$^{+4.0}_{-3.7}$ $\times$ 10$^{16}$ \\
\hline\hline                               
CH$_3$CH$_2$CN  & T (K)              &  103 $\pm$ 9   & 114 $\pm$ 11  \\
	                      &  N (cm$^{-2}$) & 3.7$^{+0.6}_{-0.5}$ $\times$ 10$^{15}$ &1.7$^{+0.3}_{-0.2}$ $\times$ 10$^{15}$ \\
\hline\hline                               
\end{tabular}\\
\end{table}%

\section{Ethyl cyanide}
\label{propio}
 
 \begin{figure}[h]
    \centering
    \includegraphics[width=0.5\textwidth]{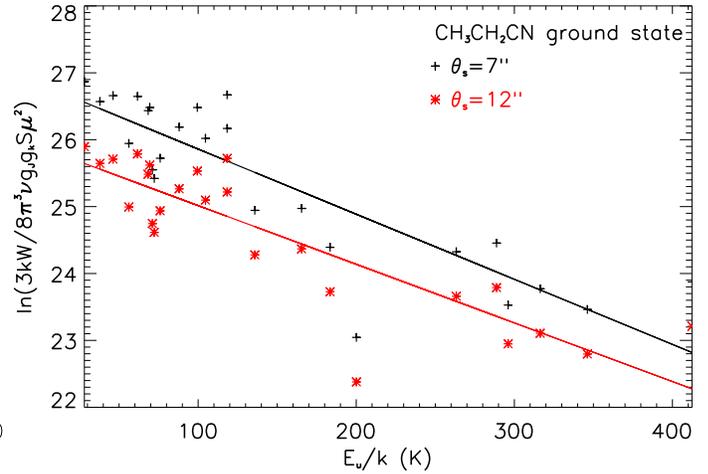}%
    \caption{Rotational diagram of the ethyl cyanide in the ground state for different source sizes. For a source size of 7\arcsec (black crosses) we find a rotational temperature of $\mathrm{T_{rot}}$ = 103 $\pm$ 9 K and a column density of N =  3.7$^{+0.6}_{-0.5}$ $\times$10$^{15}$ cm$^{-2}$; for a source size of 12\arcsec (red stars) we find $\mathrm{T_{rot}}$ = 114 $\pm$ 11 K and N =  1.3 $\times$10$^{16}$ $\times$10$^{15}$ cm$^{-2}$.}%
    \label{drpropex}
\end{figure}
 
Severals tens of lines from ethyl cyanide in the ground state are observed in the spectra. A large number of these lines are not blended and their energy covers a wide range, allowing to plot a rotational diagram (Fig. \ref{drpropex}).
Adopting a source size of 12\arcsec as for the ground state of methyl formate, we find a rotational temperature of $\mathrm{T_{rot}}$ = 114 $\pm$ 11 K and a column density N =  1.7$^{+0.3}_{-0.2}$ $\times$10$^{15}$ cm$^{-2}$. Adopting a smaller source of 7\arcsec does not change significantly the rotational temperature but changes the column density : $\mathrm{T_{rot}}$ = 103 $\pm$ 9 K and a column density N = 3.7$^{+0.6}_{-0.5}$ $\times$10$^{15}$ cm$^{-2}$  (Table \ref{tcol}).

Ethyl cyanide appears to be colder than previously found.  \citet{liu2001} adopt a temperature of 200 K and find an abundance for this molecule of 4 $\times$ 10$^{15}$ cm$^{-2}$. However their analysis is based on lines having energies lower than 113 K. If we limit ourselves to low energy transitions we find $\mathrm{T_{rot}}$ = 184 K and a column density N =  1.3 $\times$10$^{16}$ cm$^{-2}$. \citet{ikeda2001} fixed $\mathrm{T_{rot}}$ to 150 K and found a column density of N = 7 $\times$10$^{14}$ cm$^{-2}$.\\

{We have searched for transitions from the first excited bending mode (in-plane CCN bending mode) of ethyl cyanide at 206 cm$^{-1}$ (designated by $\upsilon_b$=1) and from the first torsional excited state, $\upsilon_t$=1, at 212 cm$^{-1}$.  These 2 states, together with the CCN out-of-plane bending mode at 378 cm$^{-1}$, have been studied by \citet{fukuyama1999} in the 8-200 GHz range for J and K$_a$ $\le$ than 16 and 2, respectively. Lines from the two lowest energy states $\upsilon_b$=1 and  $\upsilon_t$=1  have been detected toward SgrB2 by \citet{mehringer2004}. This paper presents the molecular theory used for the spectral analysis of new measurements of excited ethyl cyanide in the 85-400 GHz range for J and K$_a$ $\le$ than 50 and 15, respectively. However, the complete analysis has not been published yet and the prediction for the line frequencies and intensities was obtained from J. Pearson (private communication). \\

Most of the lines belonging to the excited states $\upsilon_t$=1 and $\upsilon_b$=1 are blended with other strong lines, identified or not (Table~\ref{propiovt1}). However few lines allow the column density of vibrationally excited ethyl cyanide  to be constrained (Table~\ref{propiovt1}). We have estimated the upper limit on the abundance of excited CH$_3$CH$_2$CN by comparing the emission spectrum of excited ethyl cyanide simulated with the LTE model with the observed spectra, for different temperatures and column densities (Fig.~\ref{propex}). Adopting the same temperature as the ground state (100 K) and a higher temperature of 200 K we find that the column density of vibrationally excited ethyl cyanide is 10$^{16}$cm$^{-2}$ and 5 $\times$ 10$^{15}$cm$^{-2}$, respectively.

\begin{figure*}[ht]
    \centering
    \includegraphics[width=1\textwidth]{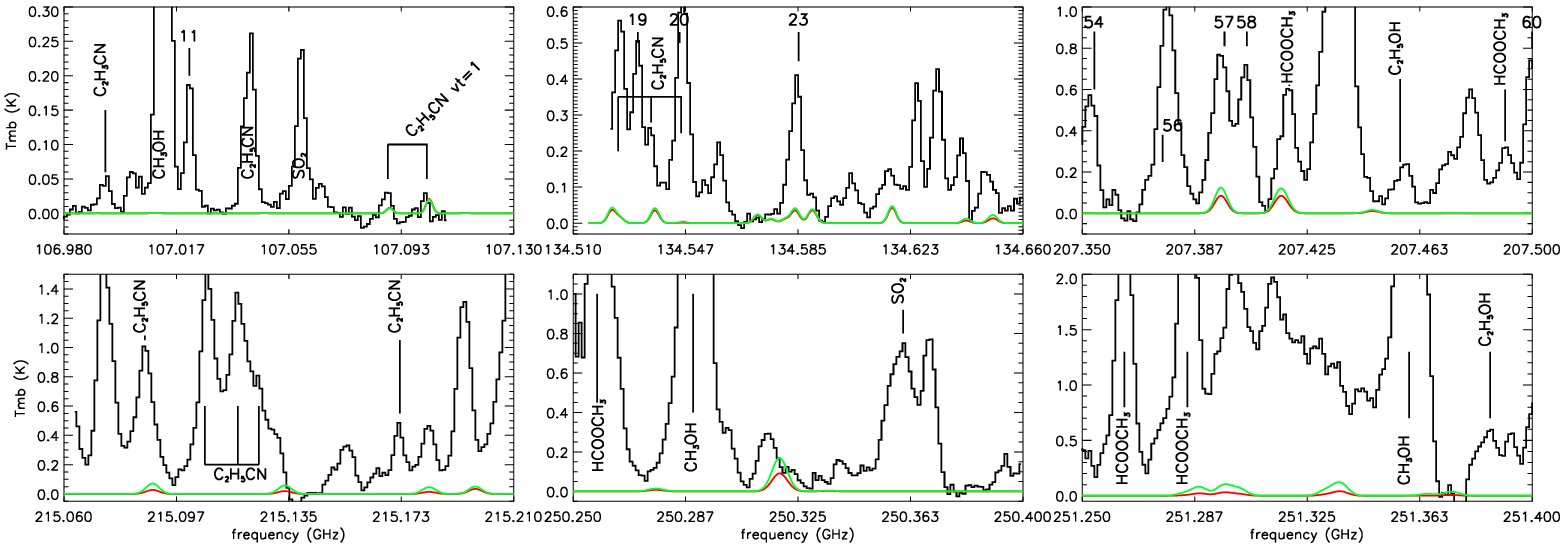}%
    \caption{Detected lines from the first two excited states $\upsilon_t$=1 and $\upsilon_b$=1 of ethyl cyanide CH$_3$CH$_2$CN. The observations (histogram like curve) are compared with LTE emission models of CH$_3$CH$_2$CN $\upsilon_t$=1 and $\upsilon_b$=1  with ${\theta}_s$ = 7", T =100 K and N = 10$^{16}$ cm$^{-2}$ (red curve) and T = 200 K and N = 5 $\times$ 10$^{15}$ cm$^{-2}$ (green curve). The numbered lines are excited methyl formate lines (see Table~\ref{mtfvt1}). See Table \ref{propiovt1} for the identification of the excited CH$_3$CH$_2$CN lines.}%
    \label{propex}
\end{figure*}

\section{Methyl carbamate and glycine}
\label{mtc}

The initial aim of the observations was to search for methyl carbamate (NH$_2$COOCH$_3$), an isomer of glycine (NH$_2$CH$_2$COOH). Methyl carbamate has a strong dipole moment and is energetically more stable than glycine, two reasons that make it a good candidate for interstellar detection. The rotational spectrum of methyl carbamate in the A torsional substate was studied in the 8-240 GHz frequency range \citep{bakri,ilyushin06}. This study was recently extended by \citet{groner07}  who measured transitions of both  A and E torsional substates up to 371 GHz. Our first analysis of the observations from 2003 was performed with a prediction of lines frequencies and intensities made for the A-transitions from the work from \citet{bakri}. We have then made a new analysis of our data from 2003 and 2006 with the predictions given by \citet{groner07}.\\

Several tens of lines from methyl carbamate were searched in spectral regions carefully chosen in order to avoid or limit confusion with spectral lines from other molecules. However we did not detect methyl carbamate in our data. Furthermore, despite everything, most of the lines are fully blended and only very few lines allow us to constrain the upper limit of the abundance of methyl carbamate in W51 e2. 
By comparing the simulated emission spectrum of methyl carbamate at different temperatures and source sizes with the observed spectrum (Fig.\ref{mcarb}) we have constrained the upper limit of methyl carbamate in this source. For a small emission region (7\arcsec) and warm gas (200 K) we found N $\le$ 5 $\times$ 10$^{14}$ cm$^{-2}$. This upper limit decreases as the rotational temperature and/or the size of the emission source decreases. We find N $\le$ 2 $\times$ 10$^{14}$ cm$^{-2}$ for $\mathrm{T_{rot}}$ = 100 K and $\mathrm{{\theta}_s}$ = 12\arcsec and  N $\le$ 8 $\times$ 10$^{13}$ cm$^{-2}$ for $\mathrm{T_{rot}}$ = 50 K and $\mathrm{{\theta}_s}$ = 30\arcsec.\\

\begin{figure*}
    \includegraphics[scale=1]{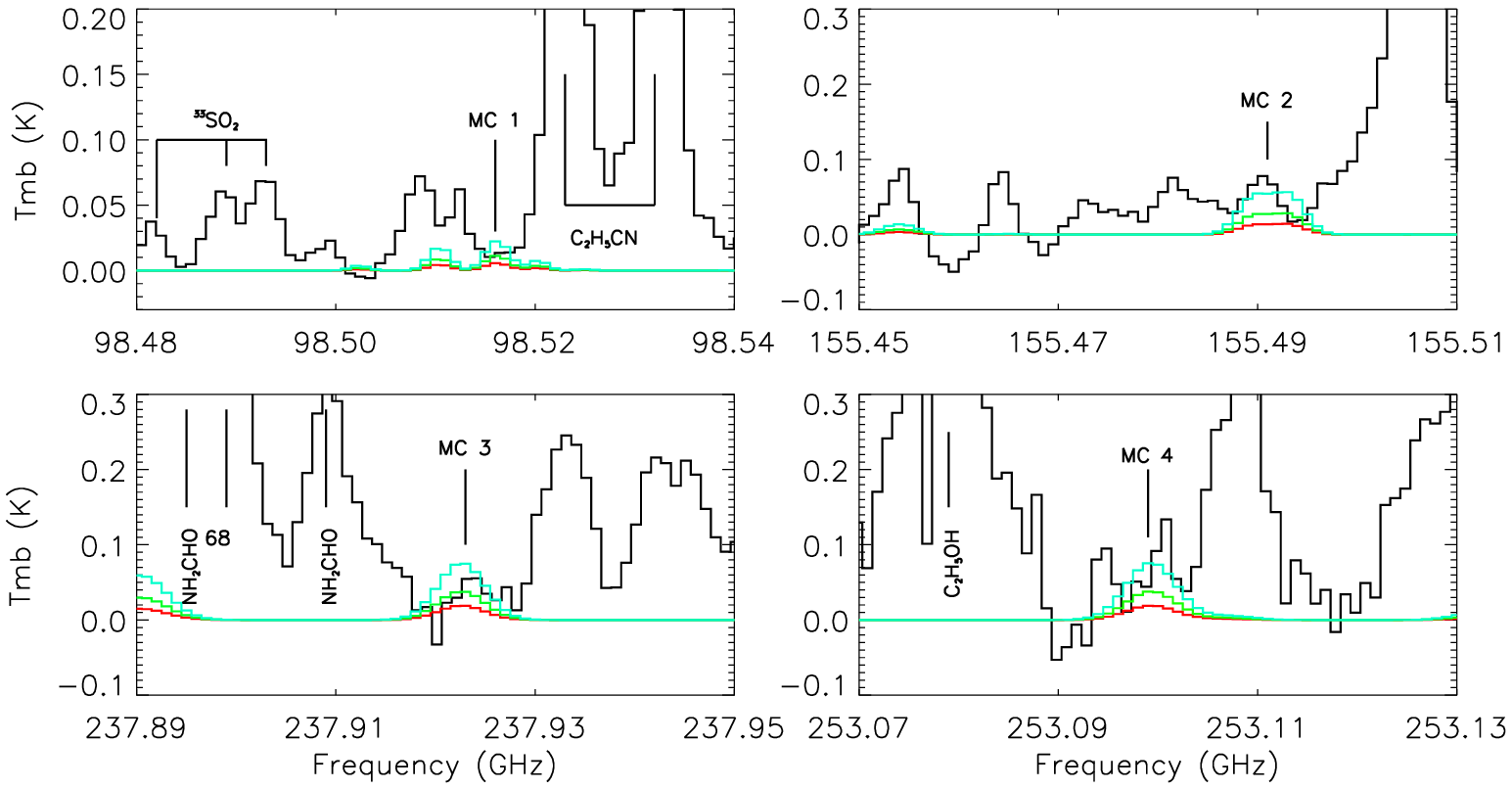}%
    \caption{Observed spectrum of W51 e2 (histogram like curve) compared with the LTE emission spectrum of methyl carbamate at different densities, for a 12\arcsec source and rotational temperature of 100 K: red curve: N = 10$^{14}$ cm$^{-2}$, green curve: N = 2 $\times$ 10$^{14}$ cm$^{-2}$, blue curve: N = 4 $\times$ 10$^{14}$ cm$^{-2}$. The methyl carbamate lines are \citep[from][]{groner07}: MC1 : $14_{2,13}-13_{1,12}$ (E) at 98515.38 GHz and $15_{1,15}-14_{0,14}$ E and A at 98516.99 GHz; MC2 :  $23_{1,22}-22_{2,21}$ at 155489.10 and 155492.68 GHz for the E and A symmetry, respectively and $23_{2,22}-22_{1,21}$ at 155490.07 and 155493.65 GHz for the E and A symmetry, respectively; MC3 : $14_{10,5}-13_{9,4}$ (A) and $14_{10,4}-13_{9,5}$ (A) (blended) at 237922.60 GHz; MC4 :  $16_{10,7}-15_{9,6}$ (A) and $16_{10,6}-15_{9,7}$ (A) at 253099.30 and  253099.37 GHz, respectively.}%
    \label{mcarb}
\end{figure*}

Numerous transitions from both conformers of glycine fall within our observed spectral ranges. Glycine is not detected in the spectra. Almost all the lines are blended and only few of them may be used to estimate roughly the upper limit of glycine in W51 e2. This upper limit varies by an order of magnitude depending on the source size, and rotational temperature.  Assuming a rotational temperature of 100 K the upper limit on the column density of glycine is 3 $\times$ 10$^{14}$ cm$^2$ for a 7\arcsec source size and 6 $\times$ 10$^{13}$ cm$^2$ for a 30\arcsec source size.

\section{Discussion}
\label{discussion}

We find that the rotational temperatures derived from the rotational diagram analysis are different from the kinetic temperature, T$_\mathrm{kin}$ =  153(21) K derived by \citet{remijan2004}. The rotational temperature of ethyl cyanide (103 - 114 K) is lower than T$_\mathrm{kin}$ independently of the adopted source size. It is higher than T$_\mathrm{kin}$ if we consider methyl formate in its ground state (176 - 199 K), lower if we consider only the excited state (104 - 131 K) and of the same order if we consider simultaneously the ground state and the first excited state of methyl formate (144 -154 K). 
All this tends to indicates that the LTE hypothesis is probably not fully valid. However, despite these possible departure from LTE, the rotational diagram method which implicitly assume that LTE is reached and that all temperature are equivalent (i.e. T$_\mathrm{exc}$ = T$_\mathrm{kin}$ = T$_\mathrm{rot}$ = T$_\mathrm{vib}$), is the only one left for us to get an estimation of the excitation temperature since a statistical analysis is ruled out by the absence of known collision rates for the studied molecules.

The relevance of the LTE assumption within each rotational level may also be evaluated by comparing the cloud density to the critical density of each vibrational state. Collisional rates are not known for molecules as complex as methyl formate or ethyl cyanide. However, it is possible to estimate roughly the critical density for these molecules by adopting the value of methanol interacting with H$_2$, of the order of a few 10$^{-11}$ cm$^3$.s$^{-1}$ \citep{pottage2004} and assuming it is the same for the ground and vibrational excited states. The Einstein coefficients of the rotational transitions are in the range 10$^{-5}$-10$^{-6}$ s$^{-1}$. The critical density is thus of the order of 10$^{5}$-10$^{6}$ cm$^{-3}$, i.e., comparable to the hydrogen density in W51 e2 estimated to be n$_\mathrm{H}$ = 5(2) $\times$ 10$^5$ cm$^{-3}$ \citep{remijan2004}. There is thus a competition between collisional and radiative excitation and it is clear that the levels are probably not all thermalized. 

The excitation mechanism populating the observed excited state of methyl formate and ethyl cyanide may be infered in a similar way. The Einstein coefficients for ro-vibrational transitions from the excited states to the ground state are of the order of a few 10$^{-1}$ s$^{-1}$, much greater than for rotational transitions within the excited state (10$^{-5}$-10$^{-6}$ s$^{-1}$). Thus, adopting the same value for the collisional rate as before (a few 10$^{-11}$ cm$^3$.s$^{-1}$), the density needed to thermalize the levels in the excited state by collisions must be greater than the critical density which is of the order of 10$^{10}$ cm$^{-3}$. The dust in W51 e2 has a temperature of about 140 K \citep{sollins2004} and thus emits efficiently at the wavelength of the excited states of both molecules. It is thus most probable that the excited states of methyl formate and ethyl cyanide are populated by radiative processes rather than by collisions.\\

We find that the rotational temperature of the gas decreases when the source size gets smaller. This is surprising since it is expected that the deepest regions of hot cores are also the warmest. The observed transitions are optically thin for the adopted source sizes of 12\arcsec and 7 \arcsec. BIMA observations of methyl formate in W51 e2 \citep{remijan2002} show that the emission region is smaller for high energy transition than for low energy transition. The size of the emission region is around 12\arcsec for the 7$_{2,5}$-6$_{2,4}$ transition at 90146 MHz corresponding to an energy of 10 cm$^{-1}$ (15 K) and around 7\arcsec for the 18$_{5,13}$-17$_{5,12}$ transition at 228629 MHz corresponding to an energy of 75 cm$^{-1}$ (108 K). More interferometric observations are needed to locate precisely the emission region of methyl formate and ethyl cyanide in the ground and excited state and understand their temperature distribution.\\

The presence of methyl formate in torsional excited state in hot cores such as W51 e2 and Orion KL \citep{koba2007} is not surprising. Methyl formate is very abundant and its torsional mode has a very low energy (188 K). By comparison, it is not surprising that ethyl cyanide is more difficult to detect: it is less abundant and its excited states lie at slightly higher energy, around 300 K. However it is clear that observations having better signal to noise, better spectral  and angular resolution, such as forthcoming observations from Herschel and ALMA, will reveal much more lines from excited states of these molecules but also of abundant large molecules possessing low-frequency vibrational states. For example dimethyl ether, CH$_3$OCH$_3$, has two torsional modes at about 203 and 242 cm$^{-1}$ and transitions from these modes should be present in the spectra of hot cores. More generally, a large number of unidentified lines reported in the spectra of hot molecular clouds should be due to transitions from abundant molecules in excited states.\\

\begin{acknowledgements}
The authors would like to thank Alexandre Faure for fruitful discussions, Jean Demaison and Isabelle Kleiner for invaluable help in the methyl formate study and John Pearson for providing us with the excited ethyl cyanide spectrum. We acknowledge all the Pico Veleta IRAM staff for their help during the observations. We thank the referee for his/her constructive comments. This work was supported by the Programme National "Physico Chimie du Milieu Interstellaire'' and by the European Research Training Network "Molecular Universe'' (MRTN-CT-2004-512302). M.C. thanks the CNRS (project CERC3) and Junta de Andalucia (project P07-FQM-03014) for financial support.
\end{acknowledgements}

\bibliographystyle{aa}
\bibliography{bibw51e2.bib}

\begin{thebibliography}{28}
\expandafter\ifx\csname natexlab\endcsname\relax\def\natexlab#1{#1}\fi

\bibitem[{{Bakri} {et~al.}(2002){Bakri}, {Demaison}, {Kleiner}, {Margul{\`e}s},
  {M{\o}llendal}, {Petitprez}, \& {Wlodarczak}}]{bakri}
{Bakri}, B., {Demaison}, J., {Kleiner}, I., {et~al.} 2002, Journal of Molecular
  Spectroscopy, 215, 312

\bibitem[{{Carvajal} {et~al.}(2007){Carvajal}, {Willaert}, {Demaison}, \&
  {Kleiner}}]{carvajal07}
{Carvajal}, M., {Willaert}, F., {Demaison}, J., \& {Kleiner}, I. 2007, Journal
  of Molecular Spectroscopy, 246, 158

\bibitem[{{Demaison} {et~al.}(1984){Demaison}, {Boucher}, \&
  {Dubrulle}}]{demaison1984}
{Demaison}, J., {Boucher}, D., \& {Dubrulle}, A. 1984, Journal of Molecular
  Spectroscopy, 102, 260

\bibitem[{{Fuchs} {et~al.}(2005){Fuchs}, {Fuchs}, {Giesen}, \&
  {Wyrowski}}]{fuchs05}
{Fuchs}, G.~W., {Fuchs}, U., {Giesen}, T.~F., \& {Wyrowski}, F. 2005, \aap,
  444, 521

\bibitem[{{Fukuyama} {et~al.}(1999){Fukuyama}, {Omori}, {Odashima}, {Takagi},
  \& {Tsunekawa}}]{fukuyama1999}
{Fukuyama}, Y., {Omori}, K., {Odashima}, H., {Takagi}, K., \& {Tsunekawa}, S.
  1999, Journal of Molecular Spectroscopy, 193, 72

\bibitem[{{Gordy} \& {Cook}(1984)}]{gordy}
{Gordy}, W. \& {Cook}, R.~L. 1984, Microwave Molecular Spectra (New York: John
  Wiley \& Sons)

\bibitem[{{Groner} {et~al.}(2007){Groner}, {Winnewisser}, {Medvedev}, {De
  Lucia}, {Herbst}, \& {Sastry}}]{groner07}
{Groner}, P., {Winnewisser}, M., {Medvedev}, I.~R., {et~al.} 2007, \apjs, 169,
  28

\bibitem[{{Hollis} {et~al.}(2006){Hollis}, {Lovas}, {Remijan}, {Jewell},
  {Ilyushin}, \& {Kleiner}}]{hollis06}
{Hollis}, J.~M., {Lovas}, F.~J., {Remijan}, A.~J., {et~al.} 2006, \apjl, 643,
  L25

\bibitem[{{Hougen} {et~al.}(1994){Hougen}, {Kleiner}, \&
  {Godefroid}}]{hougen94}
{Hougen}, J.~T., {Kleiner}, I., \& {Godefroid}, M. 1994, Journal of Molecular
  Spectroscopy, 163, 559

\bibitem[{{Ikeda} {et~al.}(2001){Ikeda}, {Ohishi}, {Nummelin}, {Dickens},
  {Bergman}, {Hjalmarson}, \& {Irvine}}]{ikeda2001}
{Ikeda}, M., {Ohishi}, M., {Nummelin}, A., {et~al.} 2001, \apj, 560, 792

\bibitem[{{Ilyushin} {et~al.}(2006){Ilyushin}, {Alekseev}, {Demaison}, \&
  {Kleiner}}]{ilyushin06}
{Ilyushin}, V., {Alekseev}, E., {Demaison}, J., \& {Kleiner}, I. 2006, Journal
  of Molecular Spectroscopy, 240, 127

\bibitem[{{Ilyushin} {et~al.}(2007){Ilyushin}, {Kleiner}, \&
  {Lovas}}]{ilyushin07}
{Ilyushin}, V., {Kleiner}, I., \& {Lovas}, F.~J. 2007, Journal of Physical and
  Chemical Reference Data

\bibitem[{{Kleiner} {et~al.}(1996){Kleiner}, {Lovas}, \&
  {Godefroid}}]{kleiner96}
{Kleiner}, I., {Lovas}, F.~J., \& {Godefroid}, M. 1996, Journal of Physical and
  Chemical Reference Data, 25, 1113

\bibitem[{{Kobayashi} {et~al.}(2007){Kobayashi}, {Ogata}, {Tsunekawa}, \&
  {Takano}}]{koba2007}
{Kobayashi}, K., {Ogata}, K., {Tsunekawa}, S., \& {Takano}, S. 2007, \apjl,
  657, L17

\bibitem[{{Liu} {et~al.}(2001){Liu}, {Mehringer}, \& {Snyder}}]{liu2001}
{Liu}, S.-Y., {Mehringer}, D.~M., \& {Snyder}, L.~E. 2001, \apj, 552, 654

\bibitem[{{Mehringer} {et~al.}(2004){Mehringer}, {Pearson}, {Keene}, \&
  {Phillips}}]{mehringer2004}
{Mehringer}, D.~M., {Pearson}, J.~C., {Keene}, J., \& {Phillips}, T.~G. 2004,
  \apj, 608, 306

\bibitem[{{Nummelin} {et~al.}(1998){Nummelin}, {Bergman}, {Hjalmarson},
  {Friberg}, {Irvine}, {Millar}, {Ohishi}, \& {Saito}}]{numm98}
{Nummelin}, A., {Bergman}, P., {Hjalmarson}, A., {et~al.} 1998, \apjs, 117, 427

\bibitem[{{Oesterling} {et~al.}(1999){Oesterling}, {Albert}, {De Lucia},
  {Sastry}, \& {Herbst}}]{oesterling1999}
{Oesterling}, L.~C., {Albert}, S., {De Lucia}, F.~C., {Sastry}, K.~V.~L.~N., \&
  {Herbst}, E. 1999, \apj, 521, 255

\bibitem[{{Ogata} {et~al.}(2004){Ogata}, {Odashima}, {Takagi}, \&
  {Tsunekawa}}]{ogata2004}
{Ogata}, K., {Odashima}, H., {Takagi}, K., \& {Tsunekawa}, S. 2004, Journal of
  Molecular Spectroscopy, 225, 14

\bibitem[{{Plummer} {et~al.}(1984){Plummer}, {Herbst}, {De Lucia}, \&
  {Blake}}]{plummer1984}
{Plummer}, G.~M., {Herbst}, E., {De Lucia}, F., \& {Blake}, G.~A. 1984, \apjs,
  55, 633

\bibitem[{{Plummer} {et~al.}(1986){Plummer}, {Herbst}, {De Lucia}, \&
  {Blake}}]{plummer1986}
{Plummer}, G.~M., {Herbst}, E., {De Lucia}, F.~C., \& {Blake}, G.~A. 1986,
  \apjs, 60, 949

\bibitem[{{Pottage} {et~al.}(2004){Pottage}, {Flower}, \&
  {Davis}}]{pottage2004}
{Pottage}, J.~T., {Flower}, D.~R., \& {Davis}, S.~L. 2004, \mnras, 352, 39

\bibitem[{{Remijan} {et~al.}(2002){Remijan}, {Snyder}, {Liu}, {Mehringer}, \&
  {Kuan}}]{remijan2002}
{Remijan}, A., {Snyder}, L.~E., {Liu}, S.-Y., {Mehringer}, D., \& {Kuan}, Y.-J.
  2002, \apj, 576, 264

\bibitem[{{Remijan} {et~al.}(2004){Remijan}, {Sutton}, {Snyder}, {Friedel},
  {Liu}, \& {Pei}}]{remijan2004}
{Remijan}, A., {Sutton}, E.~C., {Snyder}, L.~E., {et~al.} 2004, \apj, 606, 917

\bibitem[{{Snyder} {et~al.}(2005){Snyder}, {Lovas}, {Hollis}, {Friedel},
  {Jewell}, {Remijan}, {Ilyushin}, {Alekseev}, \& {Dyubko}}]{snyder2005}
{Snyder}, L.~E., {Lovas}, F.~J., {Hollis}, J.~M., {et~al.} 2005, \apj, 619, 914

\bibitem[{{Sollins} {et~al.}(2004){Sollins}, {Zhang}, \& {Ho}}]{sollins2004}
{Sollins}, P.~K., {Zhang}, Q., \& {Ho}, P.~T.~P. 2004, \apj, 606, 943

\bibitem[{{Turner}(1991)}]{turner1991}
{Turner}, B.~E. 1991, \apjs, 76, 617

\bibitem[{{Zhang} {et~al.}(1998){Zhang}, {Ho}, \& {Ohashi}}]{zhang98}
{Zhang}, Q., {Ho}, P.~T.~P., \& {Ohashi}, N. 1998, \apj, 494, 636

\end{thebibliography}

\longtab{1}{ 
\begin{longtable}{cccccccccccccc}
\caption{\label{mtfvt1lab} Laboratory measurements, calculated frequencies and line strengths for methyformate transitions in the first torsionally excited state used in the present detection.}\\
\hline\hline
 J' & K$_\mathrm{a}$' & K$_\mathrm{c}$' & P' & J" & K$_\mathrm{a}$" & K$_\mathrm{c}$" & P"  & Obs. Freq  & Calc.Freq & Calc.Unc & Obs-Calc &S${\mu}^2$ & $\mathrm{E_l}$  \\
\multicolumn{4}{c}{Upper State$^a$} & \multicolumn{4}{c}{Lower State$^a$} & (MHz)$^b$ & (MHz)$^c$ & (kHz)$^c$ & (MHz)$^c$ & (D$^2$) & (cm$^{-1}$)$^d$ \\
\hline
\endfirsthead
\caption{Laboratory measurements, calculated frequencies and line strengths for methyformate transitions in the first torsionally excited state used in the present detection. -- continued from previous page} \\
\hline \hline
 J' & K$_\mathrm{a}$' & K$_\mathrm{c}$' & P' & J" & K$_\mathrm{a}$" & K$_\mathrm{c}$" & P"  & Obs. Freq  & Calc.Freq & Calc.Unc & Obs-Calc &S${\mu}^2$ & $\mathrm{E_l}$  \\
\multicolumn{4}{c}{Upper State$^a$} & \multicolumn{4}{c}{Lower State$^a$} & (MHz)$^b$ & (MHz)$^c$ & (kHz)$^c$ & (MHz)$^c$ & (D$^2$) & (cm$^{-1}$)$^d$ \\
\hline
\endhead
\hline 
\multicolumn{14}{r}{{\it{continued on next page}}} \\ 
\endfoot
\endlastfoot
\hline
 7  	&  4 		&  4 	& -  	& 6 	&  4 		&  3 	& -  	&  85327.104 	&   85327.085 	&    11	&   0.019 	& 12.483 	& 146.8606  	\\
 7  	&  4 		&  3 	& + 	& 6 	&  4 		&  2 	& + 	&  85360.669 	&   85360.823 	&    11 	&  -0.154 	& 12.483 	& 146.8610  	\\
 7  	&  3 		&  5 	& + 	& 6 	&  3 		&  4 	& + 	&  85371.762 	&   85371.789 	&    11 	&  -0.027 	& 15.120 	& 143.6253  	\\
 7  	& -6 		&  2 	&   	&  6	&  -6 	&  1 	&   	&                	&   85456.630 	&    19 	&            	&   4.944	& 155.5393  	\\
 7  	&  4 		&  3 	&   	&  6 	&  4  	& 2  	&   	&  85506.175 	&   85506.179	&    16 	&  -0.004 	& 12.571 	& 146.6097  	\\  
 7  	& -5 		&  3 	&   	&  6 	& -5 		&  2 	&    	&  85553.365 	&   85553.288 	&    17 	&   0.077 	&  9.123 	& 150.4139   	\\
10 	&  2 		&  9 	& - 	& 10 & 1 		& 10 & + 	&  85664.038 	&   85663.943 	&    20 	&  0.095 	&  1.403 	& 151.8343     	\\
11 	&  4 		&  7 	& + 	& 11 & 3 		& 8 	& -  	&  85727.753 	&   85727.728 	&    18 	&  0.025 	&  3.078 	& 162.4132   	\\
  7 	& -4 		&  4 	&    	&  6 	& -4 		&  3	&   	&  85743.967 	&   85743.894 	&    16 	&  0.073 	& 12.532 	& 146.2356     	\\
  7 	&  2 		&  5 	& + 	&  6	& 2 		& 4 	& + 	&  88998.368 	&   88998.415 	&    12 	& -0.047 	& 17.049 	& 141.4958      	\\
  7 	&  2 		&  5 	&    	&  6 	& 2 		& 4 	&  	&  89140.383 	&   89140.366 	&    18 	&  0.017 	& 17.051 	& 141.0232    	\\
25 	&  5 		& 20 & - 	& 25 & 4 		& 21 & + 	&                    	&  107021.644 	&   361 	&         	&  8.289  	& 274.8357    	\\
  9 	& -2 		&  8 	&    	&  8 	& -2 		&  7  & 	&  107022.162 	&  107022.092 	&    13 	&   0.070 	& 22.628  	& 146.6924    	\\    
10 	&  1 		&  9 	& -  	&  9  &  2 		&  8 	& - 	&  107472.351 	&  107472.387 	&    15 	&  -0.036	&   2.144  & 150.7587   	\\
  9 	& -3 		&  7	&    	&   8 & -3  	& 6  	&    	&  111005.617 	&  111005.582 	&    19 	&   0.035 	& 20.800  	& 149.1777     	\\
  9 	&  1 		&  8 	&    	&   8 &  1 		&  7  &    	&  111094.105 	&  111094.056 	&    13 	&   0.049	& 23.222  	& 146.0900    	\\
12 	&  0 		& 12 & + 	& 11 &  0 		& 11 & + 	&  131536.624 	&  131536.653  &   12 	&  -0.029	& 31.375 	& 155.8502   	\\  
12 	&  0 		& 12 &    	& 11 &  0 		& 11 &    	&  131612.344 	&  131612.301 	&    13 	&   0.043 	& 31.558 	& 155.3072    	\\
12 	&  1 		& 12 & + 	& 11 &  0 		& 11 & + 	&  131764.316 	&  131764.341 	&    12 	&  -0.025 	&   4.870	& 155.8502    	\\
11 	&  5 		&  7 	& +  & 10 &  5 		&  6 	& + 	&  134531.846  &  134531.833 	&    13 	&   0.013 	& 23.125 	& 164.8530     	\\
27 	&  7 		& 20 & -  	& 27 &  6 		& 21 & + 	&                    	&  134545.609 	&   582 	& 	   	&   8.581 	& 304.5077     	\\
11 	& -7 		& 5  	&  	& 10 & -7 		&  4 	&   	&  134545.615 	&  134545.516 	&    27 	&   0.099 	& 17.431 	& 175.4588    	\\ 
35 	& 8 		& 27 & + 	& 35 &  7 		& 28 & -  	&               	&  134553.290  &  912 	&            	& 12.622 	& 416.6235     	\\
11 	& 5 		& 6 	& -  	& 10 &  5 		&  5 	& -	&  134585.070 	&  134585.141 	&    13  	&  -0.071 	& 23.125 	& 164.8541     	\\
11 	& -3	 	&  9	&  	& 10 & -3 		&  8 	& 	&  134713.629	&  134713.526 	&    16 	&   0.103 	& 26.708 	& 156.9827 	\\ 
11 	& 5 		&  6 	&  	& 10 &  5 		&  5  & 	&  134739.630 	&  134739.629 	&    19 	&   0.001	& 23.251 	& 164.7142     	\\
12 	& 2 		& 11 & - 	& 11 &  2 		& 10 & - 	&  140020.525 	&  140020.564 	&    12 	&  -0.039 	&  30.616	& 158.9957     	\\
11 	& 2 		& 9 	& + 	& 10 &  2 		&  8	& + 	&  140166.667 	&  140166.713 	&    12 	&  -0.046	&  28.103 	& 155.9781     	\\  
12 	& -2 		& 11 &  	& 11 & -2 		& 10 & 	&  140324.728 	&  140324.699 	&    12 	&   0.029 	&  30.787 	& 158.5211    	\\
13 	& -1 		& 13 &  	& 12 & -1 		& 12 & 	&  142032.334 	&  142032.293 	&    14 	&   0.041 	&  34.222 	& 159.7047     	\\
13  	& 0 		& 13 & + 	& 12 &  0 		& 12 & + 	&  142052.800 	&  142052.735 	&    14 	&   0.065 	&  34.029 	& 160.2378     	\\
13  	& 0 		& 13 &   	& 12 &  0 		& 12 & 	&  142125.416 	&  142125.410 	&    14 	&   0.006 	&  34.224 	& 159.6973     	\\
13  	& 1 		& 13 & + 	& 12 &  0 		& 12 & +	&  142185.220 	&  142185.198 	&    14 	&   0.022 	&    5.339 & 160.2378     	\\
15  	& 0 		& 15 & +  & 14 &  1 		& 14 & + 	&  163042.398 	&  163042.248 	&    20 	&   0.100  &    6.267 & 170.0683   	\\
15  	& 1 		& 15 & +  & 14 &  1 		& 14 & + 	&  163086.032 	&  163085.873 	&    20 	&   0.159 	&  39.344 	& 170.0683     	\\
15  	& 0 		& 15 &   	& 14 & -1 		& 14 & 	&  163113.094 	&  163113.156 	&    18	&  -0.062 	&   6.073 	& 169.5325     	\\
15  	& 0 		& 15 & + 	& 14 &  0 		& 14 & + 	&  163118.722 	&  163118.574 	&    20 	&   0.148 	&  39.345 	& 170.0657     	\\
14  	& 1 		& 13 &   	& 13 &  1 		& 12 & 	&  163142.587 	&  163142.607 	&    14	&  -0.020 	&  36.185 	& 168.1362     	\\
15  	& -1 		& 15 &  	& 14 & -1 		& 14 & 	&  163154.325 	&  163154.404 	&    18 	&  -0.079 	&  39.561 	& 169.5325     	\\
15  	& 0 		& 15 &  	& 14 &  0 		& 14 & 	&  163185.864 	&  163185.922 	&    18 	&  -0.058 	&  39.561 	& 169.5300     	\\
15  	& -1 		& 15 &  	& 14 &  0 		& 14 & 	&  163227.097 	&  163227.169 	&    18 	&  -0.072 	&    6.074 & 169.5300     	\\
16  	& 0 		& 16 & +  & 15 &  1 		& 15 & +	&  173616.616 	&  173616.407 	&    25 	&   0.209 	&    6.727 & 175.5082   	\\
16  	& 1 		& 16 & +  & 15 &  1 		& 15 & + 	&  173641.411 	&  173641.171 	&    25 	&   0.240 	&  42.007 & 175.5082   	\\
16  	& 0 		& 16 & +  & 15 &  0 		& 15 & + 	&  173660.281 	&  173660.032 	&    25 	&   0.249 	&  42.007 & 175.5068   	\\
16  	& 0 		& 16 &   	& 15 & -1 		& 15 & 	&  173683.479 	&  173683.597 	&    23 	&  -0.118 	&   6.525 	& 174.9747   	\\
24  	& 3 		& 21 &  - 	& 24 &  2 		& 22 & + 	&                    	&  173704.470 	&   241 	&            	&   4.324 	& 254.0651     	\\
16  	& -1 		& 16 &   	& 15 & -1 		& 15 & 	&  173706.683 	&  173706.807 	&    23 	&  -0.124 	&  42.232 	& 174.9747   	\\
16  	& 0 		& 16 &   	& 15 &  0 		& 15 & 	&  173724.731 	&  173724.845 	&    23 	&  -0.114 	&  42.232 	& 174.9733   	\\
16  	& -1 		& 16 &   	& 15 &  0 		& 15 & 	&  173747.990 	&  173748.055 	&    23 	&  -0.065 	&    6.525 & 174.9733   	\\
16  	& 5 		& 11 &   	& 15 &  5 		& 10 & 	&  198384.885 	&  198384.966 	&    32 	&  -0.081 	&  38.205 	& 191.3885     	\\
26  	& 8 		& 19 &-  	& 26 &  7 		& 20 & + 	&                    	&  198429.332 	&   290 	&             &   6.921 	& 297.4256   	\\
16  	& 5 		& 11 &-  	& 15 &  5 		& 10 & - 	&                    	&  198539.350 	&    30    	&           	& 38.279 	& 191.5148      	\\ 
16  	& -5 		& 12 &  	& 15 & -5 		& 11 &  	&  198578.563 	&  198578.589  &   33  	&  -0.026 	& 38.090 	& 191.0593   	\\ 
19  	& 3 		& 16 &-  	& 18 &  4 		& 15 & -	&            		&  207295.959 	&  118 	&             &   2.804 	& 208.1331   	\\
17  	& 12 	&  5 	&+  	& 16 & 12 	&  4	& + 	&        		&  207354.058 	&   101 	&	    	& 12.640 	& 252.9573      	\\
17  	& 12 	&  6 	&-  	& 16 & 12 	&  5 	& - 	&          		&  207354.058 	&   101 	&       	&  12.640 & 252.9573     	\\
17 	& -13	&  5 	&   	& 16 & -13 	&  4 	&  	&        		&  207376.777 	&   145 	&          	&  18.783 & 264.5661      	\\   
17 	&  9  	& 8  	&    	& 16 &  9 		&  7	&  	&           		&  207397.312 	&    68 	&             &  32.511 & 223.9160    	\\
17 	& 11 	&  7 	& + 	& 16 & 11 	&  6	& + 	&        		&  207404.993 	&    81 	&           	&  26.248 & 242.2767      	\\
17 	& 11 	&  6 	& -  	& 16 & 11 	&  5	& - 	&         		&  207404.993 	&   81 	&             &  26.248 & 242.2767      	\\
17 	& 10 	&  8 	& -  	& 16 & 10 	&  7	& - 	&         		&  207500.297 	&    66 	&             &  29.526 & 232.5324     	\\
17 	& 10		&  7 	& + 	& 16 & 10 	&  6	& +	&        		&  207500.297 	&    66 	& 	        &  29.526 & 232.5324    	\\
19 	&  2		& 18 & -  	& 18 &  1 		& 17	& - 	&         		&  215130.469 	&    47 	&     	        &  6.562 	& 199.2115    	\\
18	&  5		& 13 & - 	& 17 &  5 		& 12	& - 	&          		&  225648.010 	&    57 	& 	        &  44.072 & 205.2072     	\\
19 	& 2 		& 17 & + 	& 18	&  2 		& 16	& +	&        		&  225702.569 	&    46 	&         	&  48.659 & 203.7419     	\\
 6  	& 6  		& 1 	& - 	&  5 	&  5 		&  0 	& -  	&            		&  225727.540 	&    30 	&             &   2.592 	& 148.5921     	\\
 6  	& 6  		& 0 	& + 	& 5 	&  5 		& 1 	& + 	&            		&  225727.552 	&    30 	&             &   2.592 	& 148.5921   	\\
18 	&  5 		& 13 &  	& 17 &  5 		& 12 &  	&           		&  225756.154 	&    54 	&             &  43.428 	& 205.0713     	\\
 7  	& 6 		& 1 	& + 	&  6 	&  5 		&  2 	& + 	&            		&  237899.077 	&    27 	&           	&   2.594 	& 151.0254    	\\
20 	& 3  		& 18 & + 	& 19 &  2 		& 17 & + 	&        		&  237969.273 	&    58 	&             &   5.371 	& 211.2706    	\\
37 	&  7 		& 31 & + 	& 37 &  6 		& 32 & -  	&        		&  238027.934 	&  1642 	&            	&   8.537 	& 434.5487    	\\
20 	&-14  	& 7  	&   	& 19 & -14 	&  6 	& 	&          		&  244000.464 	&   195 	&             & 27.134 	& 299.1664    	\\
20 	& 15  	& 5 	& - 	& 19 & 15 	&  4 	& - 	&         		&  244048.806 	&   207 	&           	& 23.017 	& 312.5791    	\\
20 	& 15  	& 6 	& + 	& 19 & 15 	&  5 	& +  &       		&  244048.806 	&   207 	&             & 23.017	& 312.5791    	\\
19 	&  4  	& 15 & + 	& 18 &  4 		& 14 & + 	&       		&  244066.113 	&    79 	&             & 48.355	& 209.5678    	\\
27 	&  1  	& 26 & -  	& 27 &  0 		& 27 & +  &      		&  244066.575 	&   470 	&             &  1.456 	& 268.2307    	\\
20 	& 14  	& 6 	& +  & 19 & 14 		&  5 	& + 	&       		&  244073.956 	&   178	&             & 20.522 	& 299.0954    	\\
20 	& 14  	& 7 	& -  	& 19 & 14 	&  6 	& - 	&         		&  244073.956 	&   178 	&             & 20.522 	& 299.0954    	\\
20 	& 10  	& 10 &  	& 19 & 10 	&  9 	& 	&           		&  244112.669 	&   108 	& 	       	&  39.852 	& 254.7516    	\\
20 	& 13  	& 7 	& -  	& 19 & 13 	&  6 	& - 	&         		&  244119.960 	&   154 	&             &  24.670 	& 286.5471    	\\
20 	& 13  	& 8 	& + 	& 19 & 13 	&  7 	& + 	&        		&  244119.960 	&   154 	&             &  24.670 	& 286.5471    	\\
20 	& 12  	& 8 	& + 	& 19 & 12 	&  7 	& + 	&        		&  244198.512 	&   134 	&             & 34.020 	& 274.9352    	\\
20 	& 12  	& 9 	& -  	& 19 & 12 	&  8 	&  - 	&        		&  244198.512 	&   134 	&            	&  34.020 	& 274.9352    	\\
20 	& -13 	&  8 	&   	& 19 & -13 	&  7 	& 	&          		&  244207.619 	&   159 	&            	& 30.739 	& 286.5459    	\\
\hline\hline
\end{longtable}
\small{
\noindent $^a$Upper and lower state transitions quantum numbers. The rotational quantum number J, and the asymmetric rotor labels K$_\mathrm{a}$ and K$_\mathrm{c}$ are identified for each energy level. For the A symmetry species, P is the parity quantum number, for the E species, P is not defined, instead, the K$_\mathrm{a}$ label has a signed value \citep{hougen94, ilyushin07}. Note that this information has been suppressed in Table~\ref{mtfvt1}, otherwise the quantum number labeling is the same.\\
\noindent $^b$Observed laboratory frequencies in MHz from \citet{ogata2004}, the experimental uncertainty is 50 kHz for all measured lines.\\
\noindent $^c$Calculated frequencies, uncertainties and obs-calc values are from \citep{carvajal07} for the measured lines, the predicted frequencies and uncertainties for unmeasured lines is from Kleiner et al. (private communication).\\
\noindent $^d$Lower state energy in cm$^{-1}$, relative to the J=K=0 A species levels, set as zero energy.\\
}}

\longtab{2}{ 
\begin{longtable}{lcrrrrrrl}
\caption{\label{mtfvt1} Detected transitions of the first torsionally excited state of HCOOCH$_3$ in W51 e2.}\\
\hline\hline
& Transition$^a$ &  S${\mu}^2$ & $\mathrm{E_l}$ & Frequency  & Obs. frequency$^b$ &$\int \mathrm{T_{mb}}\Delta$v$^c$ & T$\mathrm{_{mb}}$  & Comment$^d$\\
&                &  (D$^2$)       & (cm$^{-1}$)       & (MHz)        & (MHz)               &  (K.km.s$^{-1}$)     &  (mK)                               &        \\
\hline
\endfirsthead
\caption{Detected transitions of HCOOCH$_3$ ${\upsilon}_t$=1 in W51 e2 -- continued from previous page} \\
\hline \hline
& Transition$^a$ &  S${\mu}^2$ & $\mathrm{E_l}$ & Frequency  & Obs. frequency$^b$ &$\int \mathrm{T_{mb}}\Delta$v$^c$ & T$\mathrm{_{mb}}$  & Comment$^d$\\
&                &  (D$^2$)       & (cm$^{-1}$)       & (MHz)        & (MHz)               &  (K.km.s$^{-1}$)     &  (mK)                               &        \\
\hline
\endhead
\hline 
\multicolumn{9}{r}{{\it{continued on next page}}} \\ 
\endfoot
\endlastfoot
\hline
1(8) 		& 7(4,4) - 6(4,3) A     	& 12.5  	& 146.86	&  85327.104 	&  85326.864 	& 1.22	& 87 	& 	\\
2(5)		& 7(4,3) - 6(4,2) A     	& 12.5  	& 146.86 	&  85360.669 	&  85359.685 	& 0.40	& 81 	&  	\\
3 		& 7(3,5) - 6(3,4) A     	& 15.1  	& 143.62 	&  85371.762 	&  85371.871 	& 0.67 	& 102 	&   	\\
4 		& 7(6,2) - 6(6,1) E     	& 4.94  	& 155.53 	&  85456.630 	&  85456.067 	& 14.33	& 1133 	& CH$_3$CCH	 \\
5(2)		& 7(4,3) - 6(4,2) E     	& 12.6  	& 146.60 	&  85506.175 	&  85506.053 	& 0.57 	& 84 	&  	\\
6 		& 7(5,3) - 6(5,2) E    		& 9.12  	& 150.40 	&  85553.365 	&  85553.361 	& 0.54 	& 78 	&  	\\
7 		& 10(2,9) - 10(1,10) A   	& 1.40  	& 151.83 	&  85664.038 	&  85664.621 	& 1.06 	& 61  	& C$_2$H$_3$CN \\
8(1) 		& 7(4,4) - 6(4,3) E     	& 12.5  	& 146.22 	&  85743.967 	&  85743.915 	& 0.79 	& 102 	&  	\\
9(10)		& 7(2,5) - 6(2,4) A   		& 17.0 	& 141.50 	&  88998.368 	&  88997.692 	& 0.77 	& 83  	&	\\
10(9) 	& 7(2,5) - 6(2,4) E 		& 17.1  	& 141.01 	&  89140.383 	&  89140.161 	& 1.20 	& 120  	&	\\
11 		& 25(5,20) - 25(4,21) A 	& 8.29  	& 274.84 	& 107021.644 	& 107021.862 	& 1.99	& 186   	&	\\
12 		& 9(2,8) - 8(2,7) E     	& 22.6  	& 146.68 	& 107022.162 	& 	-	 	&    - 	&  -	 	& 	\\
13 		& 10(1,9) - 9(2,8) A    	& 2.14  	& 150.76 	& 107472.351 	& 107472.400 	&  0.27 	&  45  	& 	\\
14 		& 9(3,7) - 8(3,6) E     	& 20.8  	& 149.16 	& 111005.617 	& 111005.832 	&  2.67	&  261  	&	\\
15 		& 9(1,8) - 8(1,7) E     	& 23.2  	& 146.08 	& 111094.105 	& 111094.010 	&  2.27	&  244  	& 	\\
16(17) 	& 12(0,12) - 11(0,11) A 	& 31.4  	& 155.85 	& 131536.624 	& 131536.949 	&  3.53	&  407 	&  	\\
17(16)	& 12(0,12) - 11(0,11) E 	& 31.6  	& 155.29 	& 131612.344 	& 131612.094 	&  3.97	&  410 	&   	\\
18 		& 12(1,12) - 11(0,11) A 	& 4.87  	& 155.85 	& 131764.316 	& 131765.100  	& 1.70	&  100  	& S$^{18}$O	\\
19 		& 11(5,7) - 10(5,6) A   	& 23.1  	& 164.85 	& 134531.846 	& 134531.411 	&  3.44	&  478 	& SO$_2$ $\upsilon_2$=1\\
20 		& 27(7,20) - 27(6,21) A 	& 8.58  	& 304.51 	& 134545.609 	& 134546.431 	& 7.95	&  630  	& C$_3$H$_2$CN + $^{33}$SO$_2$\\
21 		& 11(7,5) - 10(7,4) E   	& 17.4  	& 175.45 	& 134545.615 	& 	-	 	&  -		&  -   	& - \\
22 		& 35(8,27) - 35(7,28) A 	& 12.6  	& 416.62 	& 134553.290 	& 134553.647	&  0.71	&  991 	& $^{33}$SO$_2$ \\
23(25) 	& 11(5,6), - 10(5,5) A  	& 23.1  	& 164.85 	& 134585.070 	& 134584.263 	&  4.68	&  382  	& C$_3$H$_2$CN$\upsilon_t$,$\upsilon_b$=1 and ?\\
24 		& 11(3,9) - 10(3,8) E   	& 26.7  	& 156.97 	& 134713.629 	& 134713.062 	&  3.59 	&  456  	&  	\\
25(23) 	& 11(5,6) - 10(5,5) E   	& 23.3  	& 164.70 	& 134739.630 	& 134739.110 	&  3.55	&  386  	&  	\\
26(28) 	& 12(2,11) - 11(2,10) A 	& 30.6  	& 159.00 	& 140020.525 	& 140020.145 	&  2.96 	&  472  	&  	\\
27 		& 11(2,9) - 10(2,8) A   	& 28.1  	& 155.98 	& 140166.667 	& 140166.502 	&  3.42 	&  528  	& 	\\
28(26) 	& 12(2,11) - 11(2,10) E 	& 30.8  	& 158.50 	& 140324.728 	& 140324.321 	&  7.66	&  470 	&	\\
29 		& 13(1,13) - 12(1,12) E 	& 34.2  	& 159.69 	& 142032.334 	& 142031.811 	&  4.52 	&  586  	&  	\\
30(31) 	& 13(0,13) - 12(0,12) A 	& 34.0  	& 160.24 	& 142052.800 	& 142052.867 	&  5.95 	&  724  	&  	\\
31(30) 	& 13(0,13) - 12(0,12) E 	& 34.2  	& 159.68 	& 142125.416 	& 142125.086 	&  4.01 	&  559  	&  	\\
32 		& 13(1,13) - 12(0,12) A 	& 5.34  	& 160.24 	& 142185.220 	& 142184.202 	&  0.81	&  121  	&   	\\
33(35) 	& 15(0,15) - 14(1,14) A 	& 6.27  	& 170.07 	& 163042.398 	& 163042.050 	&  2.32 	&  363  	& 	\\
34(38) 	& 15(1,15) - 14(1,14) A 	& 39.3  	& 170.07 	& 163086.032 	& 163085.494 	&  6.39 	&  884  	& HCOOCH$_3$ $\upsilon_t$=0\\
35(33) 	& 15(0,15) - 14(1,14) E 	& 6.07  	& 169.52 	& 163113.094 	& 163113.656 	&  5.11 	&  549  	& SO$_2$\\
36(39) 	& 15(0,15) - 14(0,14) A 	& 39.3  	& 169.52 	& 163118.722 	& 163118.285 	&  19.40	&   2184  	& SO$_2$\\
37 		& 14(1,13) - 13(1,12) E 	& 36.2  	& 168.12 	& 163142.587 	& 163141.991 	&  7.69 	&  967  	& 	\\
38(34) 	& 15(1,15) - 14(1,14) E 	& 39.6  	& 162.52 	& 163154.325 	& 163153.604 	&  8.99 	&  855  	&  	\\
39(36) 	& 15(0,15) - 14(0,14) E 	& 39.6  	& 169.52 	& 163185.864 	& 163185.586 	&  5.15	&  679  	& 	\\
40 		& 15(1,15) - 14(0,14) E 	& 6.07  	& 169.52 	& 163227.097 	& 163227.290 	& 4.23 	&  448  	&  	\\
41(44) 	& 16(0,16) - 15(1,15) A 	& 6.73  	& 175.51 	& 173616.616 	& 173617.000   & 1.80  	&  177  	&  	\\
42(46) 	& 16(1,16) - 15(1,15) A 	& 42.0  	& 175.51 	& 173641.411 	& 173636.978 	& 15.14 	&  1288  	& HCOOCH$_3$ $\upsilon_t$=0 \\
43(47) 	& 16(0,16) - 15(0,15) A 	& 42.0  	& 175.51 	& 173660.281 	& 173659.384 	& 4.12  	&  380  	& 	\\
44(41) 	& 16(0,16) - 15(1,15) E 	& 6.50  	& 175.51	& 173683.479 	& 173186.987	&  17.8	&  5300	&CH$_3$CHO \\ 
45 		& 24(3,21) - 24(2,22) A 	& 4.32  	& 254.06 	& 173704.470 	& 173706.183 	&  9.03 	&  644  	& 	\\
46(42) 	& 16(1,16) - 15(1,15) E 	& 42.2  	& 174.96 	& 173706.683 	& 173706.183 	&  -		&  - 		& 	\\
47(43) 	& 16(0,16) - 15(0,15) E 	& 42.2  	& 174.97	& 173724.731 	& 173724.558	&  2.02 	&  479 	& 	\\
48 		& 16(1,16) - 15(0,15) E 	& 6.6  	& 174.97	& 173747.990 	& 173748.150	&  3.94 	&  264  	& 	\\
49(51) 	& 16(5,11) - 15(5,10) E 	& 38.2 	& 191.37 	& 198384.885 	& 198384.516 	&   5.23	&  467  	& 	\\
50 		& 26(8,19) - 26(7,20) A 	& 6.92  	& 297.43 	& 198429.332 	& 198428.202 	&   1.67 	&  231  	&  c-C$_2$H$_4$ \\
51(49) 	& 16(5,11) - 15(5,10) A 	& 38.3  	& 191.52 	& 198539.350 	& 198541.500 	&  19.44	&  1400  	&  HCOOCH$_3$ $\upsilon_t$=0\\
52 		& 16(5,12) - 15(5,11) E 	& 38.1  	& 191.05 	& 198578.563 	& 198577.928 	&   3.20	&  484 	& 	\\
53 		& 19(3,16) - 18(4,15) A 	& 2.8   	& 208.13 	& 207295.959 	& 207296.361 	&   5.16	&   159  	&  CH$_3$CH$_2$CN\\
54 		& 17(12,5) - 16(12,4) A 	& 21.2  	& 252.96 	& 207354.058 	& 207352.197 	&   5.57 	&  557  	& 	\\
55 		& 17(12,6) - 16(12,5) A 	& 21.2  	& 252.96	& 	-		& -		 	&    	-	&   -	 	& 	\\
56	 	& 17(13,5) - 16(13,4) E 	& 18.8  	& 264.55 	& 207376.777 	& 207378.506 	&   11.46	&   995 	&  CH$_2$NH\\
57 		& 17(9,8) - 16(9,7) E   	& 32.5  	& 223.90 	& 207397.312 	& 207396.134 	&  8.62	&  758 	&  	\\
58 		& 17(11,6) - 16(11,5) A 	& 26.2  	& 242.28 	& 207404.993 	& 207404.612 	&  5.67  	&  690  	& 	\\
59 		& 17(11,7) - 16(11,6) A 	& 26.2 	& 242.28	&	-		& 	-	 	&  -		&  - 		&  	\\
60 		& 17(10,8) - 16(10,7) A 	& 29.5  	& 232.53 	& 207500.297 	& 207499.521 	&  4.8  	&  897 	&  	\\
61 		& 17(10,7) - 16(10,6) A 	& 29.5  	& 232.53	& 	-		& 	-		&  - 		&  - 		&  	\\
62 		& 19(2,18) - 18(1,17) A 	& 6.56  	& 199.21 	& 215130.469 	& 215130.828 	&  2.59  	&  394 	& CH$_3$CH$_2$CN\\
63(67) 	& 18(5,13) - 17(5,12) A 	& 44.1  	& 205.21 	& 225648.010 	& 225647.974 	& 5.16   	& 1103 	& HCOOCH$_3$ $\upsilon_t$=0\\
64 		& 19(2,17) - 18(2,16) A 	& 48.7  	& 203.74 	& 225702.569 	& 225698.559   & 184  	& 1400   	& H$_2$CO  \\
65 		& 6(6,1) - 5(5,0) A     	& 2.59  	& 148.59 	& 225727.540 	& 225726.732   &  2.55	& 372  	& CH$_2$DCN \\
66 		& 6(6,0) - 5(5,1) A     	& 2.59  	& 148.59 	& 225727.552 	&  	-	   	&  	-	& -	  	& - \\
67(63) 	& 18(5,13) - 17(5,12) E	& 43.4  	& 205.06 	& 225756.154 	& 225755.178	&  7.3	&  927  	& 	\\
68 		& 7(6,1) - 6(5,2) A   		& 25.6  	& 229.84	& 237899.08 	& 237896.263 	& 15.91	& 1.667  	& NH$_2$CHO\\
69 		& 20(3,18) - 19(2,17) A 	& 5.37  	& 211.27 	& 237969.273 	& 237969.938 	&  12.80	&  1.186   & unidentified species\\
70 		& 37(7,31) - 37(6,32) A 	& 8.54  	& 434.55 	& 238027.934 	& 238026.742 	&  1.12	&  122 	&   	\\
71 		& 20(14,7) - 19(14,6) E 	& 27.1  	& 299.15 	& 244000.464 	& 243999.803 	&  7.82	& 596 	&   	\\
72 		& 20(15,6) - 19(15,5) A 	& 23.3  	& 312.58 	& 244048.806 	& 244048.230   & 90.33	&  10.84	& H$_2$CS \\
73 		& 20(15,5) - 19(15,4) A 	& 23.3  	& 312.58 	& 	-		& 	-		& 	-	&    -		&	-	 \\
74 		& 19(4,15) - 18(4,14) A 	& 18.4  	& 209.57 	& 244066.113 	& 244066.720 	&  5.49	&  730 	&   	\\
75 		& 20(14,7) - 19(14,6) A 	& 14.2  	& 299.10 	& 244073.956 	& 244073.440 	&  3.54	&  314 	& unidentified species 	\\
76 		& 20(14,6) - 19(14,5) A 	& 14.2  	& 299.10 	& 	-	 	& 	-	 	&  	-	&  -	 	&   	\\
77 		& 20(10,10) - 19(10,9) E 	& 39.9  	& 254.74 	& 244112.669 	& 244112.254 	& 2.65 	&  677 	&   	\\
78(82) 	& 20(13,8) - 19(13,7) A 	& 21.0  	& 286.55 	& 244119.960 	& 244119.708 	& 2.95 	& 886	&   	\\
79 		& 20(13,7) - 19(13,6) A 	& 21.0  	& 286.55 	& 	-		&	-		& - 	 	& -		&   	\\
80 		& 20(12,8) - 19(12,7) A 	& 34.0  	& 274.94 	& 244198.512 	& 244198.232 	& 7.43 	& 1107  	&  	\\
81 		& 20(12,9) - 19(12,8) A 	& 34.0  	& 274.94 	& 	-		&	-		&   -	  	& -	 	&  	\\
82(78) 	& 20(13,8) - 19(13,7) E 	& 30.7  	& 286.54 	& 244207.619 	& 244206.113 	& 7.33 	& 893 	& unidentified species  \\
\hline\hline
\end{longtable}
\small{
\noindent $^a$The notation for the quantum numbers is the same as Table~\ref{mtfvt1lab} except that it does not contain information on the parity of the level, i.e. the signs present in Table~\ref{mtfvt1lab} (associated to $\mathrm{K_a}$ for the E species and in a separate column for the A species) have been suppressed.\\
\noindent $^b$Observed frequencies for a systemic velocity of V$_{lsr}$ = 55.3 km.s$^{-1}$. \\
\noindent $^c$The line width is of the order of 7-8 km.s$^{-1}$. \\
\noindent $^d$This column indicates the molecules blended with the detected HCOOCH$_3$ $\upsilon_t$=1 transitions. \\
The dashes indicates that the value is the same as the one in the previous line.
}}

\longtab{3}{ 
\begin{longtable}{lcrrrrrrl}
\caption{\label{propiovt1} Detected transitions of excited CH$_3$CH$_2$CN ($\upsilon_b$ = 1 and $\upsilon_t$ = 1) in W51 e2.}\\
\hline\hline
Transition & State &  S${\mu}^2$ & $\mathrm{E_l}$ & Frequency  & Obs. frequency$^a$ &$\int \mathrm{T_{mb}}\Delta$v$^b$ & T$\mathrm{_{mb}}$  & Comment\\
                &  	      & (D$^2$)        & (cm$^{-1}$)        & (MHz)          & (MHz)                       &  (K.km.s$^{-1}$)                                         &  (mK)                        &        \\
\hline
\endfirsthead
\caption{Detected transitions of excited CH$_3$CH$_2$CN ($\upsilon_b$ = 1 and $\upsilon_t$ = 1) in W51 e2 -- continued from previous page} \\
\hline \hline
Transition & State &  S${\mu}^2$ & $\mathrm{E_l}$ & Frequency  & Obs. frequency$^a$ &$\int \mathrm{T_{mb}}\Delta$v$^b$ & T$\mathrm{_{mb}}$  & Comment\\
                &  	      & (D$^2$)        & (cm$^{-1}$)        & (MHz)          & (MHz)                       &  (K.km.s$^{-1}$)                                         &  (mK)                        &        \\
\hline
\endhead
\hline 
\multicolumn{9}{r}{{\it{continued on next page}}} \\ 
\endfoot
\endlastfoot
\hline
$11_{8,4} - 10_{8,3}$ 	&	$\mathrm{\upsilon_t=1 E}$	&    116.0 &	281.68	&	98463.46	&   98462.15	&	0.82	&	62	&		\\
$11_{8,3}-	10_{8,2}$		&	$\mathrm{\upsilon_t=1 E}$	&    - 	&	-		&	98464.02	&	-		&	-	&	-	&		\\
$11_{8,4}-	10_{8,2}$		&	$\mathrm{\upsilon_t=1 A}$	&    -		&	-		&	98464.27	&	-		&	-	&	-	&		\\
$11_{8,3}-	10_{8,3}$		&	$\mathrm{\upsilon_t=1 A}$	&    -		&	-		&	98464.32	&	-		&	-	&	-	&		\\
$11_{9,3}-	10_{9,2}$		&	$\mathrm{\upsilon_t=1 E}$	&	79.2	&	295.40	&	98538.53	&	98537.9	&	(c)	& $\le$ 15	&  CH$_3$CH$_2$CN	\\
$11_{9,2}-	10_{9,1}$		&	$\mathrm{\upsilon_t=1 E}$	&	-	&	-		&	98538.84	&	-		&	-	&	-	&	-	\\
$11_{9,3}-	10_{9,1}$		&	$\mathrm{\upsilon_t=1 A}$	&	-	&	-		&	98539.10	&	-		&	-	&	-	&	-	\\
$11_{9,2}-	10_{9,2}$		&	$\mathrm{\upsilon_t=1 A}$	&	-	&	-		&	98539.11	&	-		&	-	&	-	&	-	\\
$11_{7,4}-	10_{7,3}$		&	$\mathrm{\upsilon_b=1 E}$	&    156.3	&	258.45	&	98556.61	&	98557.5	&	(c)	& $\le$ 50	&	(c)	\\
$11_{7,5}-	10_{7,4}$		&	$\mathrm{\upsilon_b=1 E}$	&    -		&	-		&	98556.88	&	-		&	-	&	-	&	-	\\
$11_{7,5}-	10_{7,3}$		&	$\mathrm{\upsilon_b=1 A}$	&    -		&	-		&	98557.12	&	-		&	-	&	-	&	-	\\
$11_{7,4}-	10_{7,4}$		&	$\mathrm{\upsilon_b=1 A}$	&    -		&	-		&	98557.27	&	-		&	-	&	-	&	-	\\
$11_{6,6}-	10_{6,5}$		&	$\mathrm{\upsilon_b=1 E}$	&    188.1	&	248.92	&	98617.87	&	98619.1	&	2.2	&	105	&	(d)	\\
$11_{6,5}-	10_{6,4}$		&	$\mathrm{\upsilon_b=1 E}$	&    -		&	-		&	98617.91	&	-		&	-	&	-	&	-	\\
$11_{6,6}-	10_{6,4}$		&	$\mathrm{\upsilon_b=1 A}$	&    -		&	-		&	98618.09	&	-		&	-	&	-	&	-	\\
$11_{6,5}-	10_{6,5}$		&	$\mathrm{\upsilon_b=1 A}$	&    -		&	-		&	98618.19	&	-		&	-	&	-	&	-	\\
$11_{4,8}-	10_{4,7}$		&	$\mathrm{\upsilon_t=1 A}$	&    237.0	&	242.51	&	98619.45	&	-		&	-	&	-	&	-	\\
$11_{5,7}-	10_{5,5}$		&	$\mathrm{\upsilon_t=1 A}$	&    199.0	&	249.94	&	98620.23	&	-		&	-	&	-	&	-	\\
$11_{5,6}-	10_{5,6}$		&	$\mathrm{\upsilon_t=1 A}$	&    -		&	-		&	98620.33	&	-		&	-	&	-	&	-	\\
$11_{4,8}-	10_{4,7}$		&	$\mathrm{\upsilon_t=1 E}$	&    237.0	&	242.51	&	98620.38	&	-		&	-	&	-	&	-	\\
$11_{5,7}-	10_{5,6}$		&	$\mathrm{\upsilon_t=1 E}$	&    212.8	&	249.94	&	98620.41	&	-		&	-	&	-	&	-	\\
$11_{4,7}-	10_{4,6}$		&	$\mathrm{\upsilon_t=1 A}$	&    237.0	&	242.51	&	98620.80	&	-		&	-	&	-	&	-	\\
$11_{4,7}-	10_{4,6}$		&	$\mathrm{\upsilon_t=1 E}$	&    -		&	-		&	98620.82	&	-		&	-	&	-	&	-	\\
$11_{5,6}-	10_{5,5}$		&	$\mathrm{\upsilon_t=1 E}$	&    212.8	&	249.94	&	98621.02	&	-		&	-	&	-	&	-	\\
$11_{4,8}-	10_{4,7}$		&	$\mathrm{\upsilon_b=1 A}$	&    240.4	&	234.40	&	98735.64	&	98736.0	&	0.38	&	31	&		\\
$11_{4,7}-	10_{4,6}$		&	$\mathrm{\upsilon_b=1 E}$	&    233.2	&	-		&	98735.77	&	-		&	-	&	-	&		\\
$11_{4,8}-	10_{4,7}$		&	$\mathrm{\upsilon_b=1 E}$	&    -		&	234.40	&	98738.68	&	98738.8	&	0.57	&	75	&		\\
$11_{4,7}-	10_{4,6}$		&	$\mathrm{\upsilon_b=1 A}$	&    240.4	&	-		&	98739.06	&	-		&	-	&	-	&		\\
$12_{1,12	}-11_{1,11	}$	&	$\mathrm{\upsilon_b=1 E}$	&    297.2	&	229.03	&	107088.63	&	107088.5	&	0.52	&	50	&	C$_2$H$_5$OH ?	\\
$12_{2,11	}-11_{2,10	}$	&	$\mathrm{\upsilon_b=1 A}$	&    -		&	229.03	&	107088.81	&	-		&	-	&	-	&	-	\\
$12_{2,11	}-11_{2,10	}$	&	$\mathrm{\upsilon_t=1 A}$	&    294.5	&	235.76	&	107101.61	&	107101.2	&	0.34	&	37	&		\\
$12_{1,11	}-11_{1,10	}$	&	$\mathrm{\upsilon_t=1 E}$	&    -		&	235.76	&	107102.25	&	-		&	-	&	-	&		\\
$12_{8,4}-	11_{8,3}$		&	$\mathrm{\upsilon_b=1 E}$	&    165.6	&	272.74	&	107251.81	&	107252.3	&   0.53	&	40	&		\\
$12_{8,5}-	11_{8,4}$		&	$\mathrm{\upsilon_b=1 E}$	&    -		&	-		&	107252.50	&	-		&	-	&	-	&		\\
$12_{8,5}-	11_{8,3}$		&	$\mathrm{\upsilon_b=1 A}$	&    -		&	-		&	107252.56	&	-		&	-	&	-	&		\\
$12_{8,4}-	11_{8,4}$		&	$\mathrm{\upsilon_b=1 A}$	&    -		&	-		&	107252.93	&	-		&	-	&	-	&		\\
$12_{8,5}-	11_{8,4}$		&	$\mathrm{\upsilon_t=1 E}$	&    148.1	&	284.97	&	107420.75	&	107420.9  & 0.80	&	40	&	(d)	\\
$12_{8,4}-	11_{8,3}$		&	$\mathrm{\upsilon_t=1 E}$	&    -		&	-		&	107421.34	&	-		&	-	&	-	&	-	\\
$12_{8,5}-	11_{8,3}$		&	$\mathrm{\upsilon_t=1 A}$	&    -		&	-		&	107421.59	&	-		&	-	&	-	&	-	\\
$12_{8,4}-	11_{8,4}$		&	$\mathrm{\upsilon_t=1 A}$	&    -		&	-		&	107421.66	&	-		&	-	&	-	&	-	\\
$15_{4,12	}-14_{4,11	}$	&	$\mathrm{\upsilon_t=1 A}$	&    333.4	&	257.47	&	134571.52	&	134571.2	&  (c)	& $\le$ 10	&	(c)	\\
$15_{3,12	}-14_{3,11	}$	&	$\mathrm{\upsilon_t=1 E}$	&    329.8	&	257.47	&	134576.01	&	-		&	-	&	-	&	-	\\
$15_{3,13	}-14_{3,12	}$	&	$\mathrm{\upsilon_t=1 A}$	&    349.5	&	251.67	&	134589.47	&	134590.2	&	0.46	&	60	& HCOOCH$_3 \upsilon_t=1$ 	\\
$15_{2,13	}-14_{2,12	}$	&	$\mathrm{\upsilon_t=1 E}$	&    - 	&	-		&	134590.45	&	-		&	-	&	-	&	-	\\
$15_{6,10	}-14_{6,9}	$	&	$\mathrm{\upsilon_b=1 E}$	&    318.4	&	255.87	&	134616.20	&	134616.1	&	3.23	&	135	&	(d)	\\
$15_{6,9}-	14_{6,8}$		&	$\mathrm{\upsilon_b=1 E}$	&    -		&	-		&	134616.38	&	-		&	-	&	-	&	-	\\
$15_{5,11	}-14_{5,10	}$	&	$\mathrm{\upsilon_b=1 A}$	&    297.7	&	255.87	&	134616.44	&	-		&	-	&	-	&	-	\\
$15_{5,10	}-14_{5,9}$	&	$\mathrm{\upsilon_b=1 A}$	&   -		&	-		&	134616.58	&	-		&	-	&	-	&	-	\\
$15_{11,5	}-14_{11,4	}$	&	$\mathrm{\upsilon_b=1 E}$	&    140.1	&	326.77	&	134640.95	&	134640.4	&	1.93	&	257	&		\\
$15_{11,5	}-14_{11,3	}$	&	$\mathrm{\upsilon_b=1 A}$	&    -		&	-		&	134641.10	&	-		&	-	&	-	&		\\
$15_{11,4	}-14_{11,3	}$	&	$\mathrm{\upsilon_b=1 E}$	&   -		&	-		&	134641.11	&	-		&	-	&	-	&		\\
$15_{11,4	}-14_{11,4	}$	&	$\mathrm{\upsilon_b=1 A}$	&    -		&	-		&	134641.12	&	-		&	-	&	-	&		\\
$15_{12,4	}-14_{12,3	}$	&	$\mathrm{\upsilon_b=1 E}$	&    104.7	&	343.90	&	134648.79	&	134648.1	&	2.37	&	132	&	(d)	\\
$15_{12,3	}-14_{12,2	}$	&	$\mathrm{\upsilon_b=1 E}$	&    -		&	-		&	134648.90	&	-		&	-	&	-	&	-	\\
$15_{12,4	}-14_{12,2	}$	&	$\mathrm{\upsilon_b=1 A}$	&    -		&	-		&	134648.98	&	-		&	-	&	-	&	-	\\
$15_{12,3	}-14_{12,3	}$	&	$\mathrm{\upsilon_b=1 A}$	&    - 	&	-		&	134649.00	&	-		&	-	&	-	&	-	\\
$15_{10,6	}-14_{10,5	}$	&	$\mathrm{\upsilon_b=1 E}$	&    175.0	&	311.15	&	134650.29	&	-		&	-	&	-	&	-	\\
$15_{10,6	}-14_{10,4	}$	&	$\mathrm{\upsilon_b=1 A}$	&    -		&	-		&	134650.34	&	-		&	-	&	-	&	-	\\
$15_{10,5	}-14_{10,5	}$	&	$\mathrm{\upsilon_b=1 A}$	&   -		&	-		&	134650.34	&	-		&	-	&	-	&	-	\\
$15_{10,5	}-14_{10,4	}$	&	$\mathrm{\upsilon_b=1 E}$	&    -		&	-		&	134650.49	&	-		&	-	&	-	&	-	\\
$15_{13,3	}-14_{13,2	}$	&	$\mathrm{\upsilon_b=1 E}$	&	69.2	&	362.56	&	134666.30	&	131667.6	&	(c)	& $\le$ 10	&	(c)	\\
$15_{13,2	}-14_{13,1	}$	&	$\mathrm{\upsilon_b=1 E}$	&	- 	&	-		&	134666.33	&	-		&	-	&	-	&	-	\\
$15_{13,3	}-14_{13,1	}$	&	$\mathrm{\upsilon_b=1 A}$	&	-	&	-		&	134666.49	&	-		&	-	&	-	&	-	\\
$15_{13,2	}-14_{13,2	}$	&	$\mathrm{\upsilon_b=1 A}$	&	-	&	-		&	134666.50	&	-		&	-	&	-	&	-	\\
$15_{9,7}-	14_{9,5}$		&	$\mathrm{\upsilon_b=1 A}$	&    209.4	&	297.07	&	134703.14	&	134702.7	&	0.98	&	103	&		\\
$15_{9,6}-	14_{9,6}$		&	$\mathrm{\upsilon_b=1 A}$	&    -		&	-		&	134703.26	&	-		&	-	&	-	&		\\
$15_{9,7}-	14_{9,6}$		&	$\mathrm{\upsilon_b=1 E}$	&    -		&	-		&	134703.45	&	-		&	-	&	-	&		\\
$15_{9,6}-	14_{9,5}$		&	$\mathrm{\upsilon_b=1 E}$	&    -		&	-		&	134703.54	&	-		&	-	&	-	&		\\
$23_{0,23	}-22_{0,22	}$	&	$\mathrm{\upsilon_b=1 E}$	&    514.8	&	280.29	&	198463.40	&	198465.5	&	4.17	&	348	&	unidentified species	\\
$23_{0,23	}-22_{0,22	}$	&	$\mathrm{\upsilon_b=1 A}$	&    -		&	-		&	198463.42	&	-		&	-	&	-	&	-	\\
$23_{4,19}-22_{4,18}$	&	$\mathrm{\upsilon_b=1 E}$	&    485.2	&	293.79	&   	207396.15	&	207397.5	&	8.58	&	758	&	HCOOCH$_3 \upsilon_t=1$	\\
$23_{4,20}-22_{4,19}$	&	$\mathrm{\upsilon_b=1 A}$	&    -		&	293.79	&   	207396.27	&	-		&	-	&	-	&	-	\\
$23_{1,22}-22_{1,21}$	&	$\mathrm{\upsilon_t=1 A}$	&   502.5	&	290.47	&   	207415.76	&	207240.2	&	6.69	&	568	&	HCOOCH$_3 \upsilon_t=1$	\\
$23_{1,23}-22_{1,22}$	&	$\mathrm{\upsilon_t=1 E}$	&   -		&	290.47	&   	207417.59&		-	&	-	&	-	&	-	\\
$24_{0,24}-23_{1,23}$	&	$\mathrm{\upsilon_t=1 E}$	&   188.2	&	293.36	&   	207446.75&	207446.0	&	(c)	& $\le$ 15	&	C$_2$H$_5$OH	\\
$24_{0,24}-23_{1,23}$	&	$\mathrm{\upsilon_b=1 A}$	&  186.9	&	293.36	&   	207447.53&		-	&	-	&	-	&	-	\\
$24_{11,14}-23_{11,13}$	&	$\mathrm{\upsilon_t=1 E}$	&    325.5	&	393.52	&	215133.09	&	215132.5	&	(c)	&$\le$ 380	& unknown species	\\
$24_{11,13}-23_{11,12}$	&	$\mathrm{\upsilon_t=1 E}$	&    -		&	-		&	215133.41	&	-		&	-	&	-	&	-	\\
$24_{11,14}-23_{11,12}$	&	$\mathrm{\upsilon_t=1 A}$	&    -		&	-		&	215134.13	&	-		&	-	&	-	&	-	\\
$24_{11,13}-23_{11,13}$	&	$\mathrm{\upsilon_t=1 A}$	&    -		&	-		&	215134.19	&	-		&	-	&	-	&	-	\\
$25_{4,21	}-24_{4,20}$	&	$\mathrm{\upsilon_b=1 E}$	&    511.8	&	307.94	&	225715.53	&	225714.7	&	(c)	& $\le$ 228&	(c) 	\\
$25_{4,22	}-24_{4,21}$	&	$\mathrm{\upsilon_b=1 A}$	&   -		&	-		&	225715.61	&	-		&		&		&		\\
$27_{4,23	}-26_{4,22}$	&	$\mathrm{\upsilon_b=1 E}$	&    534.4	&	323.30	&	244142.10	&	244139.3	&	3.94	& 	288	&		\\
$27_{4,24	}-26_{4,23}$	&	$\mathrm{\upsilon_b=1 A}$	&    -		&	-		&	244142.13	&	-		&	-	&	-	&		\\
$28_{1,27	}-27_{1,26}$	&	$\mathrm{\upsilon_t=1 A}$	&    539.3	&	327.94	&	250318.38	&	250316.2	&	2.7	&	298	&		\\
$28_{2,27	}-27_{2,26}$	&	$\mathrm{\upsilon_t=1 E}$	&    269.5	&	327.94	&	250319.81	&	-		&	-	&	-	&		\\
$28_{2,27	}-27_{2,26}$	&	$\mathrm{\upsilon_t=1 E}$	&    -		&	-		&	250319.81	&	-		&	-	&	-	&		\\
$28_{15,13}-27_{15,12}$	&	$\mathrm{\upsilon_b=1 E}$	&    273.8	&	486.17	&	251372.86	& 	251373.7	&	(c)	& $\le$ 10	&	(c)	\\
$28_{15,14}-27_{15,13}$	&	$\mathrm{\upsilon_b=1 E}$	&   -		&	-		&	251373.06	&	-		&	-	&	-	&	-	\\
$28_{15,14}-27_{15,12}$	&	$\mathrm{\upsilon_b=1 A}$	&    -		&	-		&	251373.29	&	-		&	-	&	-	&	-	\\
$28_{15,13}-27_{15,13}$	&	$\mathrm{\upsilon_b=1 A}$	&    -		&	-		&	251373.32	&	-		&	-	&	-	&	-	\\
\hline\hline
\end{longtable}
\small{
\noindent $^a$Observed frequencies for a systemic velocity of V$_{\mathrm{lsr}}$ = 57 km/s. \\
\noindent $^b$The line width is of the order of 7-8 km.s$^{-1}$. \\
\noindent $^c$Line within the noise, a gaussian fit could not be made, constrains the abundance of excited ethyl cyanide. \\
\noindent $^d$Very broad lines, ($\Delta$v $\ge$ 15$ \mathrm{km.s^{-1}}$), due to the blending of several excited ethyl cyanide transitions and/or to the blending with an unknown species.\\
The dashes indicates that the value is the same as the one in the previous line.
}}

\end{document}